\documentclass[useAMS,usenatbib]{mn2e}

\usepackage{graphicx}

\newcommand{\kms}{km s$^{-1}$}
\newcommand{\cmN}{cm$^{-2}$}
\newcommand{\cmn}{cm$^{-3}$}

\newcommand{\lam}{$\lambda$}

\newcommand{\civ}{\mbox{C\,{\sc iv}}}

\newcommand{\nv}{\mbox{N\,{\sc v}}}

\newcommand{\ovi}{\mbox{O\,{\sc vi}}}

\newcommand{\feii}{\mbox{Fe\,{\sc ii}}}

\newcommand{\neviii}{\mbox{Ne\,{\sc viii}}}
\newcommand{\lya}{\mbox{Ly$\alpha$}}

\newcommand{\aj}{AJ} 
\newcommand{\mnras}{MNRAS} 
\newcommand{\apj}{ApJ} 
\newcommand{\apjl}{ApJ} 
\newcommand{\apjs}{ApJS} 
\newcommand{\aap}{A\&A} 
\newcommand{\araa}{ARA\&A} 

\newcommand{\pasp}{PASP} 

\newcommand{\nat}{Nat} 

\title[Extreme-Velocity Quasar Outflows]{Extreme-Velocity Quasar Outflows and the Role of X-ray Shielding}
\author[F. Hamann et al.]{F. Hamann$^{1}$\thanks{E-mail:
fhamann@ufl.edu (FH)}, G. Chartas$^{2}$, S. McGraw$^{3}$, P. Rodriguez Hidalgo$^{4}$, J. Shields$^{3}$, \newauthor 
D. Capellupo$^{1,5}$, J. Charlton$^{6}$, M. Eracleous$^{6}$\\
$^{1}$Department of Astronomy, University of Florida, Gainesville, FL 
32611-2055, USA\\
$^{2}$Department of Physics \& Astronomy, College of Charleston, Charleston, SC 29424, USA\\
$^{3}$Department of Physics \& Astronomy, Ohio University, Athens, OH 45701, USA\\
$^{4}$Department of Physics \& Astronomy, York University, Toronto, ON, M3J IP3, Canada\\
$^{5}$School of Physics \& Astronomy, Tel Aviv University, Tel Aviv, 69978, Israel\\
$^{6}$Department of Astronomy \& Astrophysics, Pennsylvania State University, University Park, PA 16802, USA\\
}

\begin{document}

\date{Accepted xxx. Received xxx}

\pagerange{\pageref{firstpage}--\pageref{lastpage}} \pubyear{2008}

\maketitle

\label{firstpage}

\begin{abstract}
Quasar accretion disk winds observed via broad absorption lines (BALs) in the UV produce strong continuous absorption in X-rays. The X-ray absorber is believed to serve critically as a radiative shield to keep the outflow ionizations low enough for radiative driving. However, previous studies have shown that ``mini-BAL" and narrow absorption line (NAL) outflows have dramatically less X-ray absorption than BALs. Here we examine X-ray and rest-frame UV spectra of 8 mini-BAL quasars with outflow speeds in the range 0.1c to 0.2c to test the hypothesis that these extreme speeds require a strong shield. We find that the X-ray absorption is weak or moderate, with neutral-equivalent column densities $N_H < {\rm few} \times 10^{22}$ \cmN , consistent with mini-BALs at lower speeds. We use photoionization models to show that the amount of shielding consistent with our data is too weak to control the outflow ionizations and, therefore, it is not important for the acceleration. Shielding in complex geometries also seems unlikely because the alleged shield would need to extinguish the ionizing far-UV flux while avoiding detection in X-rays and the near-UV.

We argue that the outflow ionizations are kept moderate, instead, by high gas densities in small clouds. If the mini-BALs form at radial distances of order $R \sim 2$ pc from the central quasar (broadly consistent with theoretical models and with the mini-BAL variabilities observed here and in previous work), and the total column densities in the mini-BAL gas are $N_H\la 10^{21}$ \cmN , then the total radial extent of outflow clouds is only $\Delta R_{clouds}\la 3\times 10^{13}$ cm in cases of no/weak shielding or $\Delta R_{clouds}\la 3\times 10^{14}$ cm behind the maximum shield allowed by our data. This implies radial filling factors $\Delta R_{clouds}/R\la 5\times 10^{-6}$ or $\la$$5\times 10^{-5}$ for the unshielded or maximally shielded cases, respectively. Compared to the transverse sizes $\ga$$8\times 10^{15}$ cm (based on measured line depths), the outflows have shapes like thin ``pancakes" viewed face-on, or they occupy larger volumes like a spray of many dense clouds with a small volume filling factor. These results favor models with magnetic confinement in magnetic disk winds. To the extent that BALs, mini-BALs, and NALs probe the same general outflow phenomenon, our result for dense substructures should apply to all three outflow types. 
\end{abstract}

\begin{keywords}
galaxies: active --- quasars: general --- quasars: absorption lines
\end{keywords}

\section{Introduction}

High-velocity outflows are an important part of the quasar phenomenon. They are often studied via broad absorption lines (BALs) in the rest-frame UV that reveal outflow speeds from a few thousand to tens of thousands of \kms\ \citep{Weymann91,Korista93,Trump06}. Some studies suggest that BAL outflows play an important role in ``feedback" to the quasar's host galaxy evolution -- contributing to galaxy-scale blowouts of gas and dust, disrupting star formation in the galaxy hosts, and regulating the growth of the central supermassive black hole \citep[SMBH,][]{Silk98,Kauffmann00,DiMatteo05,Moll07,Moe09,Dunn10,Faucher12b,Farrah12}. Nonetheless, many aspects of quasar outflows remain poorly understood, including their basic physical conditions and acceleration mechanism(s). 

BAL outflows are believed to arise from quasar accretion disks, driven out by radiation pressure \citep{Arav94,Chelouche03,Murray95,deKool97,Proga00,Proga04} or magneto-hydrodynamic or magneto-centrifugal forces \citep{Konigl94,Everett05,Fukumura10}. Radiation pressure is expected to be important in luminous quasars, e.g., at high accretion rates relative to Eddington \citep{Proga07,Everett05}. However, the intense radiation available to push the outflows can also over-ionize the gas and make it too transparent for radiative driving. 

\cite{Murray95} and \cite{Murray97} proposed to solve the over-ionization problem by noting that a highly-ionized and radiatively thick absorbing region should develop naturally at the base of BAL outflows, near the quasar's intense source of ionizing radiation. This additional absorbing medium is itself too ionized and too transparent for radiative driving, but it serves critically as a shield to block ionizing radiation and thereby allow the BAL gas behind it to reach sufficient opacities for radiative acceleration. The general picture of BAL outflows behind a thick radiative shield has become a mainstay of theoretical models \citep{Chelouche03,Proga00,Proga04,Proga07,Sim10}. Even the magneto-hydrodynamic models invoke large column densities of shielding gas to launch BAL outflows with moderate ionizations from the strong gravity environment near the central SMBH \citep{Everett05,Fukumura10}. 

The shielding hypothesis is also supported by BAL quasar observations that reveal strong X-ray absorption with typical neutral-equivalent column densities $N_H \ga 10^{23}$ \cmN\ \citep{Green96,Green01,Mathur00,Gallagher99,Gallagher02}. Moreover, sources with stronger X-ray absorption tend to have larger outflow speeds and stronger \civ\ \lam\lam 1548,1550 absorption lines \citep{Brandt00,Laor02,Gallagher06}. These results suggest that X-ray shielding is important for the development of BAL outflows and, specifically, that a thicker radiative shield leads to more efficient radiative acceleration. 

However, this picture is complicated by observations of other quasars with narrow absorption line (NAL) outflows and  so-called ``mini-BALs,'' which have smooth BAL-like profiles but velocity widths below the standard BAL threshold \citep[FWHM $\la$ 2000 \kms ,][]{Weymann91,Hamann04,Hamann12}. These narrow line outflows are more common than BALs \citep{Nestor08,Paola08,Misawa07,Simon12,Hamann12} with high speeds and degrees of ionization (typified by \civ\ and \ovi\ \lam 1032,1038 absorption) similar to BALs, but they have dramatically less X-ray absorption \citep[][and \S4.2 below]{Brandt00,Misawa08,Gibson09a,Chartas09,Chartas12,Hamann11}. 

The similar properties of BAL, NAL, and mini-BAL outflows suggest that they arise from the same general outflow phenomenon, while orientation or temporal/evolution effects might explain their important differences. In one popular orientation-based scheme, BALs form in the main part of the outflow near the accretion disk plane while NALs and mini-BALs form along sightlines at higher latitudes that (perhaps) skim the ragged edges of the BAL flow farther above the disk \citep{Ganguly01,Hamann08,Chartas09,Hamann12}. This geometry is broadly consistent with the theoretical models mentioned above \citep[also][]{Proga12} and it can explain the weak or absent X-ray absorption in NAL and mini-BAL quasars if the X-ray absorber resides primarily near the plane of the accretion disk \citep{Murray95}. 

The difficulty arises when we consider that NAL and mini-BAL outflows achieve the same high speeds and moderate degrees of ionization as BALs {\it without} the protection of a radiative shield. This suggests that the shield is not a critical feature of the winds. It appears to undermine the main premise of all current radiative acceleration models that a strong shield is essential to moderate the outflow ionizations and launch gas behind the shield to high speeds. It might also require us to abandon current models with smooth continuous flows in favor of earlier schemes that involve small dense clouds with an overall small volume filling factor \citep[see \S6 below, also][]{Hamann11}. 

In this paper, we present new X-ray and rest-frame UV observations of 7 quasars with extreme-velocity mini-BAL outflows, supplemented by archival data for a similar quasar, PG 2302+029 \citep{Jannuzi96,Sabra03}. The outflow speeds in these quasar are in the range 0.1c to 0.2c, which is 2-3 times larger than typical BALs or mini-BALs in previous X-ray studies. The extreme speeds require favorable conditions for the outflow acceleration. We aim to test whether these conditions involve a strong radiative shield. 

We might expect stronger shielding in extreme-velocity outflows because they should originate from small accretion disk radii, near the ionizing UV emission source. In particular, the maximum flow velocities, $v_{\infty}$ (at infinity), are expected to scale roughly with the gravitational escape speed, $v_{esc}$, at the launch radius, $R_1$, such that $v_{\infty}\propto v_{esc}\propto 1/\sqrt{R_1}$ \citep[][also \S4.1 below]{Murray95,Proga00,Everett05}. Therefore, the extreme outflow speeds in our mini-BAL sample, 2-3 times larger than previous studies, might correspond to launch radii that are 4-9 times smaller where the ionizing flux is 16-81 times greater. 

We will show that the X-ray absorption is weak or moderate and radiative shielding is not important in these outflows. \S2 provides information about the quasar sample. \S3 describes our new UV and X-ray observations. \S4 presents some basic analysis of the mini-BAL variabilities with constraints on the outflow locations, as well as measures of the X-ray absorption and absorber properties. \S5 describes Cloudy photoionization simulations that place quantitative constraints on the amount and importance of shielding across the full UV to X-ray spectrum. \S6 discusses the implications for quasar outflow models, and \S7 provides a brief summary. Throughout this paper, we adopt a cosmology with $H_o = 71$ \kms\ Mpc$^{-1}$, $\Omega_M = 0.27$ and $\Omega_{\Lambda}=0.73$.  

\section[]{The Mini-BAL Quasar Sample}

Our study focusses on 8 quasars known to have broad \civ\ outflow lines at extreme velocities $>$0.1$c$. Six of them were discovered by \cite{Paola08} in spectra from the Sloan Digital Sky Survey \citep[SDSS,][]{Schneider07}. The other two have been studied more extensively (e.g., J093857+412821 by \citealt{Hamann97b,Paola11}; PG 2302+029 by \citealt{Jannuzi96,Jannuzi02,Sabra03}). All 8 quasars are radio quiet\footnote{If we assume the nominal {\sl FIRST} survey upper limit of 0.45 mJy at 20 cm corresponds to 0.25 mJy at 6 cm, and we use observed fluxes at 4400 \AA\ in the SDSS spectra (crudely extrapolated over the \lya\ forest in the higher redshift sources),  then the radio-loudness parameters $R = f_{\rm 6 cm}/f_{\rm 4400A}$ \citep{Kellermann89} range from $<$0.3 for the visibly brightest source (J093857+412821) to $<$0.8 for the faintest (J090508+074151).} based on non-detections in the {\sl FIRST} radio survey \citep{Becker95}. 

Figure 1 shows the \civ\ outflow lines from the studies cited above plus own new measurements at the MDM Observatory (see  \S3.2 below). These outflow features have velocities in the range $38,880 \la v \la 56,000$ \kms\ and widths $1850 \la {\rm FWHM} \la 4240$ \kms . Some of the lines are broad enough to qualify as BALs, with FWHM $>$ 2000 \kms , except that their speeds are too high to register a balnicity index BI $>$ 0 \citep{Weymann91}. Thus they are not BALs by this formal definition and, in any case, they are narrower and weaker than the majority of bona fide BALs studied previously at lower speeds. We will refer to all of the outflow lines in our sample as ``mini-BALs" \citep[see][for more discussion]{Paola08}. 

\begin{figure*}
 \includegraphics[scale=0.87,angle=-90.0]{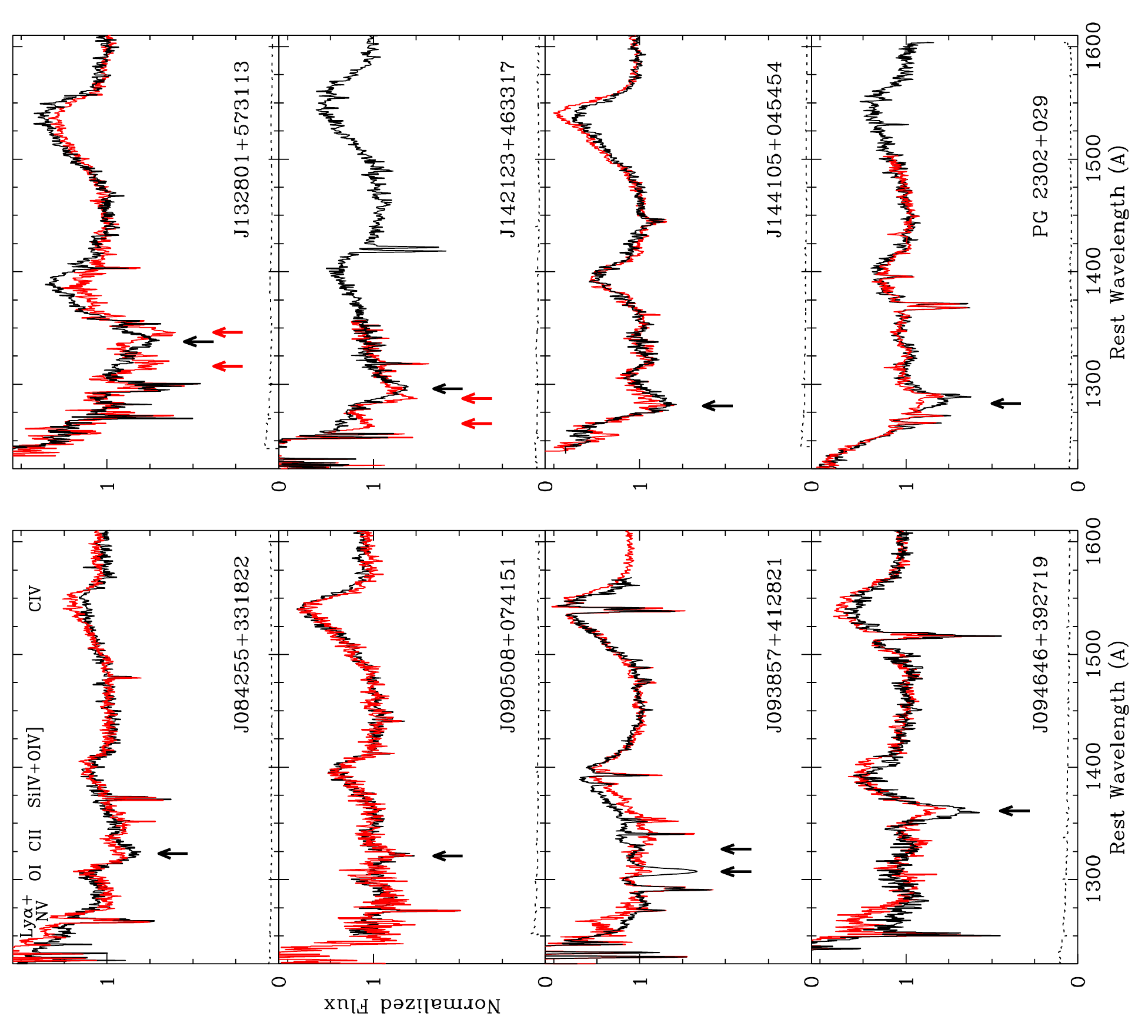}
\vspace{-10pt}
 \caption{Normalized rest-frame UV spectra of our quasar sample from two epochs. The red curves show the most recent data from the MDM Observatory except for PG~2302+029 from {\sl HST} in 1998. The black curves show earlier observations from {\sl HST} (PG~2302+029 in 1994), the Lick Observatory (J093857+412821 in 1996), or the SDSS (all others, see Table 1, \S2, and \S3.2). The black dotted curves are error spectra from the SDSS or HST, not available for other data. Significant mini-BALs in the earlier and later epochs are marked below by black and red arrows, respectively. Red arrows are not drawn if the mini-BAL disappeared or did not substantially shift its wavelength in the newer data. The locations of several broad emission lines are shown across the top in the upper left panel. }
\end{figure*}

Table 1 lists some basic data for the quasars and mini-BALs. The emission line redshifts, $z_{em}$, are from \cite{Hewett10} and the $i$ magnitudes are from the SDSS for all sources except PG2302+029, for which the redshift and $B$ magnitude are from \cite{Schmidt83}. The mini-BAL centroid velocities, $v$, and FWHMs are from \cite{Paola08}, except for PG 2302+029 from an HST spectrum obtained by \cite{Jannuzi96} in 1994.58 and for J093857+412821 (also called PG 0935+417) from a 1996.2 spectrum obtained at the Lick observatory 3.0m Shane telescope \citep[][also Figure 1]{Hamann97b,Paola11}. A second weaker \civ\ mini-BAL at $v = 45910$ \kms\ in the Lick spectrum of J093857+412821 is marked in Figure 1 but not listed in Table 1. For J090508+074151, we re-examined the SDSS spectrum measured by \cite{Paola08} and chose to use a lower, more conservative continuum placement. This yields a weaker and narrower absorption line that now appears questionable as a mini-BAL. The value of FWHM $\sim$ 740 \kms\ in Table 1 is largely due to the \civ\ doublet separation ($\sim$500 \kms ).
Columns 6--8 in Table 1 provide the observation dates for the SDSS, Chandra, and MDM data in decimal years. 

\begin{table*}
 \centering
 \begin{minipage}{140mm}
  \caption{Quasar/Mini-BAL Data and Observation Dates}
  \begin{tabular}{@{}lccccccc@{}}
  \hline
 Quasar Name& $z_{em}^a$& $i^a$ & $v^b$ & FWHM$^b$& SDSS& Chandra & MDM\\
 & & (mag) & (\kms )& (\kms )& (yr) & (yr)& (yr)\\
\hline
  J084255.61+331822.58  & 2.620& 17.43 & 44750 &    2200  & 2003.10 & 2011.01 & 2011.02, 2012.09\\
  J090508.85+074151.24   & 2.041& 17.57 & 46540 &    ~740  & 2003.92 & 2011.01 & 2011.02, 2012.09\\
  J093857.02+412821.19  & 1.957& 16.24 & 51260  &  1850  & 2003.08 & 2012.10 & 2012.07\\
  J094646.94+392719.02   & 2.206& 17.45 & 38880   &  2620  & 2003.19 & 2011.16 & 2012.08\\
  J132801.25+573113.09   & 2.074& 17.02 & 41780    & 2800  & 2003.33 & 2011.50 & 2011.31\\
  J142123.99+463317.88  & 3.378& 17.22 & 52250    & 2960  & 2003.24 & 2011.47 & 2011.32\\
  J144105.53+045454.95   & 2.068& 17.08  &  55230   & 4240  & 2001.30 & 2011.96 & 2011.32, 2012.08\\
  PG 2302+029 & 1.044 & 16.03 & 56000 & 2475 & ... & 2000.84 & ...\\
\hline
\end{tabular}
$^a$Emission line redshifts from \cite{Hewett10} and $i$ magnitudes from the SDSS, except PG2302+029 where the redshift and $B$ magnitude (not $i$) are from \cite{Schmidt83}.\\
$^b$Mini-BAL data from \cite{Paola08} adjusted for the newer redshifts in column 2, except J090508+074151 remeasured from the SDSS spectrum here (\S2), J093857+412821 measured from a 1996.2 Lick  spectrum \citep{Paola11}, and PG 2302+029 measured from a 1994.58 HST spectrum \citep{Jannuzi96}.
\end{minipage}
\end{table*}

\section[]{New Observations \& Results}

\subsection{Chandra X-Ray Observations}

We observed the first 7 quasars listed in Table 1 with the Chandra Advanced CCD Imaging Spectrometer \citep[ACIS,][]{Garmire03}. PG 2302+029 was observed previously with Chandra ACIS in November 2000 by \cite{Sabra03}. We retrieved the PG 2302+029 data from the Chandra archives, and analyzed all 8 quasars using the software package CIAO 4.4 with CALDB version 4.4.8 provided by the {\sl Chandra X-ray Center}. To construct the X-ray spectra, we extract events from circular regions with radii of 5 arcsec centered on the sources. The backgrounds are determined from events within annuli centered on the source with inner and outer radii of 7.5 arcsec and 50 arcsec, respectively. We then fit spectral models to the data restricted to events with energies between 0.3--10~keV. The spectra of J094646+392719, J132801+573113, J142123+463317, and  J144105+045454 have relatively low signal-to-noise ratios (S/N) owing to the low total counts (see below). For these quasars we binned the spectra to have at least one count per bin and performed fits using the $C$-statistic \citep{Cash79}. For J090508+074151, J093857+412821, and PG 2302+029, which have more counts and moderate S/N, we performed fits using the $\chi^{2}$ statistic. J084255+331822 was not detected by our observations. 

Table 2 lists parameters from the X-ray measurements. $t_{exp}$ is the effective exposure time after removing portions of the observations severely contaminated by background. These times are just a few percent less than the actual total exposure times. Columns 3 and 4 in Table 2 list the total X-ray counts and integrated fluxes, $F$, in the observed energy range 0.2$-$10 keV. We also measured fluxes separately in soft (0.2$-$2 keV) and hard (2.0$-$10 keV) energy bands to calculate the hardness ratio, HR (column 5), which we define as the ratio of hard/soft fluxes. $f_{2{\rm keV}}$ is the observed flux density at rest energy 2 keV. These fluxes are based on single powerlaw fits to the spectra, corrected for small amounts of Galactic absorption \citep{Dickey90,Kalberla05}. For J084255+331822, Table 2 provides 3$\sigma$ upper limits on the total counts and integrated flux equal to a 3$\sigma$ fluctuation in the background level inside a circle of radius 5 arcsec centered on the visible SDSS position. 

\begin{table*}
 \centering
 \begin{minipage}{174mm}
  \caption{UV and X-Ray Continuum Data}
  \begin{tabular}{@{}lcccccccccc@{}}
  \hline
 Quasar Name & $t_{exp}$ & Counts & $F$ & HR & ~~~$f_{2{\rm keV}}$~~ & $f_{2500{\rm A}}$& ~~$\alpha_{ox}$~~ & ~$\Delta \alpha_{ox}$~& \multispan{2}{\hfil Fit Results\hfil}\\ 
 & (ksec) & (0.2-10keV)& (0.2-10keV)& & & & & & $N_H$& $\Gamma$\\
\hline
  J084255.61+331822.58  &   30.2& $4\pm 3$   & \llap{$<\,$}6.50 &  ...  & \llap{$<\,$}3.50 & 1.88 & \llap{$<$}$-$1.89 & \llap{$<$}$-$0.16& ...& ...\\[+2pt]
  J090508.85+074151.24  & 34.1& $366\pm 21$  & 101$^{+12}_{-14}$ & 0.42  & 45.4$^{+8.5}_{-7.3}$ & 1.86 &  $-$1.44 & +0.24& 0.3$^{+0.7}_{-0.3}$& 2.37$^{+0.31}_{-0.26}$\\[+3pt]
  J093857.02+412821.19  & 14.7& $220\pm 15$  & 163$^{+25}_{-36}$  & 2.31  & 30.2$^{+7.0}_{-7.2}$ & 5.80 & $-$1.75& +0.00& 2.0$^{+1.9}_{-1.6}$& 1.69$^{+0.52}_{-0.29}$\\[+3pt]
  J094646.94+392719.02  & 27.3& $26\pm 6$   & 9.62$^{+3.5}_{-9.6}$   &  3.29  & 1.33$^{+0.8}_{-1.3}$ & 1.70 & $-$2.07& $-$0.37& 3.7$^{+11}_{-3.6}$& 1.75$^{+2.03}_{-1.15}$\\[+3pt]
  J132801.25+573113.09  & 22.4& $123\pm 12$  & 80.4$^{+13}_{-54}$    & 4.03  & 9.32$^{+3.0}_{-6.4}$ & 3.02 & $-$1.83& $-$0.11& 4.2$^{+6.6}_{-3.2}$& 1.53$^{+0.66}_{-0.55}$\\[+3pt]
  J142123.99+463317.88  & 26.6& $52\pm 8$ & 29.2$^{+6.7}_{-16}$  & 3.17  & 4.43$^{+3.2}_{-3.9}$ & 1.26 & $-$1.88& $-$0.12& 5.1$^{+8.1}_{-5.1}$& 1.50$^{+0.60}_{-0.70}$\\[+3pt]
  J144105.53+045454.95  & 21.5& $79\pm 9$ &  48.0$^{+8.1}_{-20}$  & 3.10  & 6.96$^{+2.8}_{-3.6}$ & 2.84 & $-$1.86& $-$0.15& 5.4$^{+5.1}_{-3.6}$& 1.90$^{+0.68}_{-0.56}$\\[+3pt]
  PG 2302+029 & 47.93 & $412\pm 21$ & 75.5$^{+9.1}_{-9.1}$ & 3.04& 13.0$^{+1.0}_{-1.0}$ & 15.0& $-$2.09& $-$0.40& 0.5$^{+0.2}_{-0.2}$& 1.68$^{+0.27}_{-0.18}$\\
\hline
\end{tabular}
Flux units: $F$: 10$^{-15}$ ergs cm$^{-2}$ s$^{-1}$; $f_{2{\rm keV}}$: 10$^{-15}$ ergs cm$^{-2}$ s$^{-1}$ keV$^{-1}$; $f_{2500{\rm A}}$: 10$^{-16}$ ergs cm$^{-2}$ s$^{-1}$ \AA$^{-1}$. $N_H$ units: $10^{22}$ \cmN . Uncertainties listed for the fluxes and fit results indicate 90\% $\chi^2$ confidence. Fit results for PG2302+029 are for $z_{abs}=0.56$.
\end{minipage}
\end{table*}

$f_{2500{\rm A}}$ in Table 2 is the continuum flux density measured from the SDSS spectra at rest wavelength 2500 \AA . These fluxes are corrected for Galactic extinction and for weak broad emission lines (e.g., \feii ) that might be present. For J142123+463317, the rest wavelength 2500 \AA\ lies outside the SDSS spectral coverage. We estimate its 2500 \AA\ flux by extrapolating the measured SDSS spectrum toward longer wavelengths. For PG 2302+029, we crudely estimate $f_{2500{\rm A}}$ from the $B$ magnitude listed in Table 1.

The next measured quantities in Table 2 are the two-point optical-to-X-ray power law spectral index $\alpha_{ox} = 0.384\, \log (f_{2keV}/f_{2500{\rm A}})$ \citep{Tananbaum79}, and $\Delta\alpha_{ox}=\alpha_{ox}$(observed)$\,-\, \alpha_{ox}$(predicted), which is the difference between the observed and predicted values of $\alpha_{ox}$. The predicted values derive from an empirical correlation between $\alpha_{ox}$ and the luminosity density at 2500 \AA\ in optically selected, radio-quiet quasars without BALs \citep[Eqn.2 in][]{Steffen06}. The last two columns in Table 2 give fit results described in \S4.2.2 below.

\subsection{MDM Observations}
 
We obtained new rest-frame UV spectra for 7 quasars in our sample (excepting PG 2302+029) with the OSMOS spectrograph \citep{Martini11} on the 2.4m Hiltner telescope at MDM Observatory. The observation dates (Table 1) are close in time to the Chandra measurements. We used the VPH grism with a 1.2 arcsec slit  to provide resolution $R\approx 1600$ ($\sim$190 \kms ) and wavelength coverage from roughly 3100 \AA\ to 5900 \AA . This resolution approximately matches the SDSS spectra \citep[at $R\approx 2000$, e.g., ][]{Adelman08}. The spectra cover important lines at wavelengths from at least 1200 to 1600 \AA\ in the quasar rest frames. The only exception is the high redshift source J142123+463317, for which we cover the \civ\ mini-BALs but rest wavelengths longer than $\sim$1350 \AA\ are not measured.

We reduced the MDM spectra using standard techniques with the IRAF\footnote{IRAF is distributed by the National Optical Astronomy Observatory, which is operated by the Association of Universities for Research in Astronomy (AURA) under cooperative agreement with the National Science Foundation.} software package. Relative flux calibrations were achieved using standard stars measured on the same night. Absolute fluxes are not available. 

Figure 1 shows the MDM spectra shifted to rest wavelengths (red curves) on top of the previous data described in \S2 above (black curves). Figure 1 also shows spectra of PG~2302+029 obtained with {\sl HST} using the STIS in 1998 (red curve) the FOS in 1994 (black). Mini-BALs in the earlier epochs are marked by black arrows below the black spectra. Red arrows are added only if the mini-BALs appear at significantly different wavelengths in the later epochs. The spectral resolutions are all in the range 150 to 230 \kms , which easily resolves the mini-BALs of interest here. The spectra shown in Figure 1 are normalized to unity in the quasar continuum using IRAF software with low-order polynomial fits constrained by wavelength regions not affected by strong emission or absorption lines. In most cases, the spectra are also smoothed by 2 or 3 pixel wide boxcar functions to improve the display.

For J084255+331822, J090508+074151, and J144105+045454, the MDM spectra in Figure 1 are averages from the two observing dates listed in Table 1. We inspected the individual spectra before averaging to check for variations in the mini-BALs. No significant differences were found. The averages are weighted by the noise in each spectrum (i.e., the reciprocal variance) to maximize the final signal-to-noise ratio. All other spectra in Figure 1 are from single epoch observations, as indicated in Table 1.

 \section[]{Analysis}

\subsection{Notes on Mini-BAL Variability \& Outflow Locations}

At least 5 of the 7 securely identified mini-BALs in our sample exhibit significant variability. (We exclude the questionable mini-BAL in J090508+074151 from this discussion, \S2.) The other 2 sources, J084255+331822 and J144105+045454, also appear to have mini-BAL variations but those occurrences are marginal compared to the noise and possible small changes in the broad emission lines (Figure 1). The mini-BALs in PG 2302+029 were known previously to vary between the 1994.58 HST spectrum shown in Figure 1 and a second HST spectrum obtained in 1998.98 \citep{Jannuzi02,Sabra03}. Dramatic variations are also documented in J093857+412821 \citep{Hamann97b, Narayanan04, Paola11}. In this quasar, the mini-BALs apparent in the 1996.2 Lick spectrum (Figure 1) were absent from earlier observations in 1981 \citep{Bechtold84}. They then evolved in complicated ways between at least 1993 and 2008, with changes in the strength, centroid, and overall shape of the absorption profile \citep{Paola11}. The distinct mini-BALs at 51,260 \kms\ and 45,910 \kms\ in the 1996.2 spectrum are not recognizable at other epochs. Spectra after 2003 show a shallower and much broader feature than in 1996.2, centered near $-$47,500 \kms\ ($\sim$1320 \AA\ rest). Our 2012.07 MDM spectrum (Figure 1) is consistent with no outflow absorption at all. However, the complexity of past variations suggests that the broad dip near 1350 \AA\ could still be the remnants of an outflow absorption feature. 

Variability is a common characteristic of BAL and mini-BAL outflows 
\citep{Lundgren07,Misawa07c,Gibson08,Gibson10,Capellupo11,Capellupo12,Capellupo13,Paola12,FilizAk12}. The fraction of BAL or mini-BAL quasars that exhibit line variability between two measurements separated by a few years in the rest frame has been estimated to be $\sim$60-80\%, depending on the time scale and absorber properties \citep[e.g.,][]{Paola12,Capellupo13}. For example, there are significant trends for increasing variability at higher velocities and in weaker (less deep) portions of BAL troughs \citep{Capellupo11}. Our observations are broadly consistent with these results. As such, they support the standard picture of BALs and mini-BALs probing the same general outflow phenomenon (\S1). 

Variability can provide important constraints on the outflow locations. BAL and mini-BAL changes are often attributed to flow structures crossing our lines of sight to the emission  source(s). In this situation, the variability times constrain the crossing speeds given an estimate of the emission region size. For outflow lines that absorb the quasar continuum flux, as in our sample, the relevant emission region is the UV-emitting accretion disk. \cite{Capellupo13} describe the constraints available for quasars with luminosities and predicted emitting region sizes similar to the quasars in our sample. Their Figure 15 shows that the variability times we measure, $\sim$2-3 yrs (rest), correspond to crossing speeds crudely $\ga$1000 \kms\ for simple geometries. If the crossing speeds are tied roughly to the Keplarian speeds around the central SMBH, then the radial distance of these absorbing regions should be $\la$10 pc. For J093857+412821, with much better temporal sampling, these same arguments indicate crossing speeds $\ga$5000 \kms\ and radial distances $\la$1 pc \citep{Paola11}. 

An alternative possibility is that the line variations are caused by changes in the quasar's ionizing flux. This leads to much weaker constraints on the outflow distances \citep[][and refs. therein]{Hamann97,Hamann11,Misawa07c}. It is likely that both mechanisms operate to different degrees in different sources. In many cases, the cause of the variability is not known. However, among BALs, there are numerous examples of optically thick or dramatic/complex line profile changes that strongly support the moving cloud and small distance interpretation \citep[][and Capellupo et al., in prep.]{Hamann08,Hall11,Vivek12,Capellupo12,Capellupo13}. In our mini-BAL sample, the emergence and subsequent complex profile evolution in J093857+412821 (see refs. above) is much more readily attributed to  moving clouds than changes in the ionizing flux. That result and the analogies to BAL variability lead us to favor moving clouds and distances $\la$10 pc. 

We note that some studies that use excited-state lines for density constraints argue that BALs can form at distances $\sim$1 kpc or beyond \citep{Moe09,Dunn10,Borguet13}. There is strong evidence for large distances like this in several cases of narrow low-velocity (``associated") absorption lines \citep{Hamann01}, and it does seem plausible that BALs might form across a wide range of scales. However, the studies favoring large BAL distances based on low densities do not appear to consider the effects of radiative shielding, which, for BAL quasars that are heavily absorbed in X-rays, can dramatically reduce the distances at which low-density gas can produce the observed lines \citep{Everett02}. In any case, those studies still favor small $<$1 pc distances for the variable portions of BAL troughs \citep{Moe09}, consistent with our interpretation of the mini-BALs. 

Another consideration for the outflow locations is the launch radius. Outflows with extreme velocities are expected to arise from locations with extreme gravity, such that the measured flow speeds are crudely similar to the gravitational escape speed, $v\sim v_{esc}$, at the launch site \citep{Murray95,Proga00,Everett05}. For the mini-BALs in our sample, with $v\sim 0.1$c to 0.2c, this implies launch radii of order $R \sim 0.01$ pc from a $10^9$ M$_{\odot}$ black hole. The observed mini-BALs need not form at this radius, but they might because moderate ionizations capable of \ovi\ and possibly \civ\ absorption are needed near the launch point if the flows are radiatively accelerated (\S1, also Hamann et al., in prep.). 

As a starting point for our discussions in \S5 and \S6 below, we adopt a conservative order of magnitude location for the mini-BAL gas similar to the radius of the broad emission line regions. This is consistent with the line variability, the theoretical models discussed above \citep[also][]{Murray97,Proga10}, and observational studies that link the outflows to broad emission line properties \citep{Leighly04,Richards11}. For the quasars in our sample, the H$\beta$ broad emission line radius should be near $R\sim 1$--2 pc \citep[][based on $f_{2500{\rm A}}$ in Table 1 and an assumed spectral slope $f_{\nu}\propto\nu^{-0.5}$ between 2500\AA\ and 5100\AA ]{Bentz09}. 

\subsection{X-Ray Absorption \& Comparisons to Previous Work}

Here we characterize the X-ray absorption compared to previous studies of BAL and mini-BAL quasars and to simple theoretical models that assume neutral absorbing slabs. The results for neutral absorbers are provided for illustration and comparisons to previous work. It is understood that more realistic ionized absorbers require larger total column densities to produce the same amount of X-ray absorption. We discuss ionized absorbers in \S5 below. 

\subsubsection{Results Based on HR and $\Delta\alpha_{ox}$}

Table 3 (described fully in \S5 below) gives the predicted values of $\Delta\alpha_{ox}$ and HR for neutral absorbing slabs with total column densities, $\log(N_H/{\rm cm}^{-2}) = 22$, 22.5, and 23 (named neutral22, neutral22.5, and neutral23, respectively, in the table). These calculations assume solar metal abundances with photoelectric cross sections from \cite{Morrison83}. The unabsorbed X-ray spectra are powerlaws with two different indices: $\alpha_x = -0.9$ (for $f_{\nu}\propto \nu^{\alpha_x}$), which is typical of quasars \citep{Reeves00,Just07}, and $\alpha_x = -0.5$, which provides a better match to some sources in our sample (see spectral fitting results below). The absorbed spectra for the $\alpha_x = -0.9$ case are shown by the green curves in Figure 4 (also described in \S5 below). The hardness ratios based on observed 0.2--2.0 keV and 2.0--10 keV fluxes depend on the quasar redshift. Table 3 gives predicted HR values for $z=2.0$ and $z=3.3$, representative of our sample. 

Comparing the observed and predicted values of $\Delta\alpha_{ox}$ and HR (in Tables 2 and 3, respectively) indicates that the X-ray absorption in our quasar sample is generally weak or moderate, with neutral equivalent column densities $\log(N_H/{\rm cm}^{-2}) < 22.7$. $\Delta\alpha_{ox}$ is probably a better indicator of X-ray absorption than HR, but the uncertainties for both parameters are signficant. HR can be unreliable for faint sources because the low numbers of counts in {\sl Chandra} ACIS data tend to bunch up near the boundary between ``soft" and ``hard" at 2 keV, providing little leverage to measure the spectral slope. Similarly, the measured HRs are not sensitive to the adopted energy range from 0.2 to 10 keV (we would get essentially the same results considering only 0.5 to 8 keV), but this choice can matter a lot for the theoretical results. $\Delta\alpha_{ox}$ and HR also have uncertainties due to object-to-object scatter in the emitted spectra  \citep[][and refs. therein]{Reeves00,Steffen06}. The scatter in $\alpha_{ox}$ and $\Delta\alpha_{ox}$ is known to be roughly $\pm$0.15 \citep{Just07,Steffen06,Strateva05}. $\Delta\alpha_{ox}$ can have additional uncertainties due to continuum variability if the UV and X-ray measurements are not simultaneous. This uncertainty is probably also of order $\pm$0.15 \citep{Gallagher06}. 

The non-simultaneity of the UV and X-ray measurements for PG 2302+029 might explain the discrepancy between the large negative $\Delta\alpha_{ox}$ in Table 2, crudely indicating $\log(N_H/{\rm cm}^{-2}) \sim 22.7$ (Table 3), and our spectral fits below that favor $\log(N_H/{\rm cm}^{-2}) \la 22.0$. We consider the fit results more reliable because the X-ray spectrum is well measured (with $\sim$405 counts) and the low redshift yields good coverage across the spectral energies of soft X-ray absorption. 

\cite{Gallagher06} measured $\Delta\alpha_{ox}$ and hardness ratios for 35 luminous BAL quasars at redshifts $z\sim 1.5$ to 2.9. Their distribution of $\Delta\alpha_{ox}$ values has a broad peak near $\Delta\alpha_{ox} \sim -0.6$, with $\sim$75\% of their quasars having $\Delta\alpha_{ox} < -0.4$. In contrast, the mini-BAL quasars in our sample have characteristically $\Delta\alpha_{ox}\approx -0.1$ with a sample standard deviation $\pm0.2$ \citep[see also][]{Steffen06,Just07}. \cite{Gallagher06} provide hardness ratios in the form of an X-ray photon index $\Gamma_{HR} = 1-\alpha_x$. A typical value for their BAL quasars is $\Gamma_{HR}\sim 1$, with most sources (ignoring marginal BALs and redshifts $<$2) in the range $0.6\la\Gamma_{HR}\la 1.3$. This corresponds to energy index $\alpha_x\sim 0$ with a range $+0.4\ga\alpha_x\ga -0.3$, and to hardness ratios as defined here of typically HR $\sim$ 4.4 with a range $8.9\ga {\rm HR}\ga 2.6$. \cite{Gallagher06} note that the BAL quasar hardness ratios are larger than expected from $\Delta\alpha_{ox}$ by at least a few tenths. They attribute this to significant degrees of ionization in the absorbers and/or partial covering of the background light source. Altogether, the \cite{Gallagher06} results indicate that BAL quasars have characteristically harder and much fainter X-ray spectra, indicating stronger X-ray absorption, than our extreme-velocity mini-BAL quasars. 

\cite{Gibson09a} and \cite{Wu10} obtained similar results from samples of BAL and mini-BAL quasars selected from the SDSS. Their combined sample of $\sim$60 mini-BAL quasars has an average value of $\Delta\alpha_{ox} = -0.03\pm 0.02$, which is similar to the average for non-BAL quasars, $0.00\pm 0.01$, but significantly less than their reported average for BAL quasars, $\Delta\alpha_{ox} = -0.22\pm 0.04$. We note that this average for BAL quasars is smaller than the \cite{Gallagher06} result,  probably due to a much wider distribution of BAL properties in \cite{Gibson09a} and \cite{Wu10}. In particular, the \cite{Gallagher06} sample has a preference for strong classic BALs while \cite{Gibson09a} and \cite{Wu10} use a relaxed definition of BALs\footnote{Also note that some of the ``mini-BALs'' in \cite{Gibson09a} and \cite{Wu10} might not be true broad outflow lines. They select \civ\ mini-BALs based on absorption index AI $>$ 0, which requires $\ge$1000 \kms\ of contiguous absorption at least 10\% below the continuum \citep{Trump06}. Our experience comparing higher resolution spectra to the SDSS is that features with AI $>$ 0 can turn out to be some other type of galactic or intervening absorbing gas, e.g., the ones with small velocity shifts and profiles with square sides or narrow-line substructure (see \citealt{Hamann01,Hamann08,Hamann11} and \citealt{Simon12} for examples). An expanded mini-BAL sample in \cite{Wu10}, not used for comparisons here, uses a more relaxed definition of AI that requires only $\ge$500 \kms\ of contiguous absorption $>$10\% below the continuum. Given that the \civ\ doublet separation is $\sim$500 \kms , the nominal resolution of SDSS spectra is $\sim$150 \kms , and unrelated intervening \civ\ systems often have multiple components up to $\sim$200 \kms\ apart \citep{Sargent88,Misawa07}, it seems certain that this intended outflow sample is significantly contaminated by non-outflow lines (see also the SDSS studies of narrow line absorbers by \citealt{Nestor08} and \citealt{Wild08}).} that includes absorption down to $v=0$ \citep[compared to the usual balnicity index that requires $v>3000$ \kms ,][]{Weymann91}. Nonetheless, the main result seems secure: mini-BAL quasars have weak or moderate X-ray absorption compared to the strong absorption that accompanies BALs. 

\subsubsection{X-ray Spectral Fitting}

We characterize the X-ray absorption further by fitting the X-ray spectra using the XSPEC software package. These fits again assume powerlaw emitted spectra attenuated by neutral absorbers. The absorber column densities are freely varied, along with the intrinsic spectral index and overall flux normalization. Thus the fits  ignore information provided by $\alpha_{ox}$ and $\Delta\alpha_{ox}$ discussed above. Fixed amounts of Galactic absorption appropriate for each quasar are included (\S3.1). The absorber redshifts are fixed at the quasar redshift for all sources except PG2302+029, which we discuss in more detail below. 

Figure 2 shows the resulting confidence contours for the photon index, $\Gamma$, and total neutral column density, $N_H$. The best fit values of $N_H$ and $\Gamma$ are listed in the last two columns of Table 2, with uncertainties indicating the 90\% $\chi^2$ confidence range. The uncertainties are generally large because of the small numbers of counts (Table 2). Most of the data are consistent with $N_H \sim 0$ at 90\% confidence. J144105+045454 in Figure 2 provides formally the best evidence for significant absorption in our new {\sl Chandra} observations, with $\log(N_H/{\rm cm}^{-2}) \sim 22.7$. However, this result should be viewed in comparison to the value of $\log(N_H/{\rm cm}^{-2}) \sim 22.2$ predicted by $\Delta\alpha_{ox}\sim -0.15$ (c.f., Tables 2 and 3). 

\begin{figure*}
\begin{center}
 \includegraphics[scale=0.31,angle=0.0]{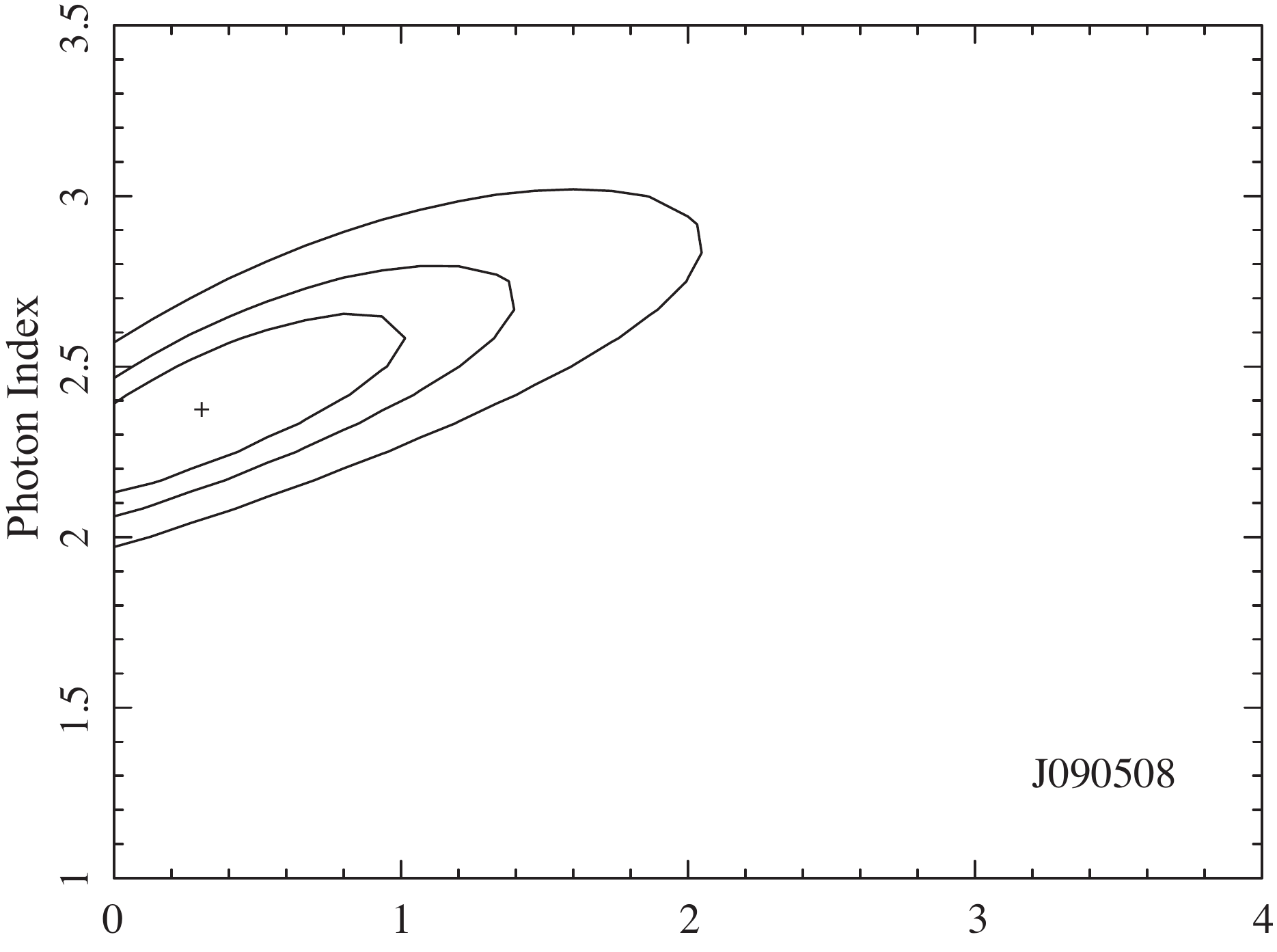}
 \includegraphics[scale=0.31,angle=0.0]{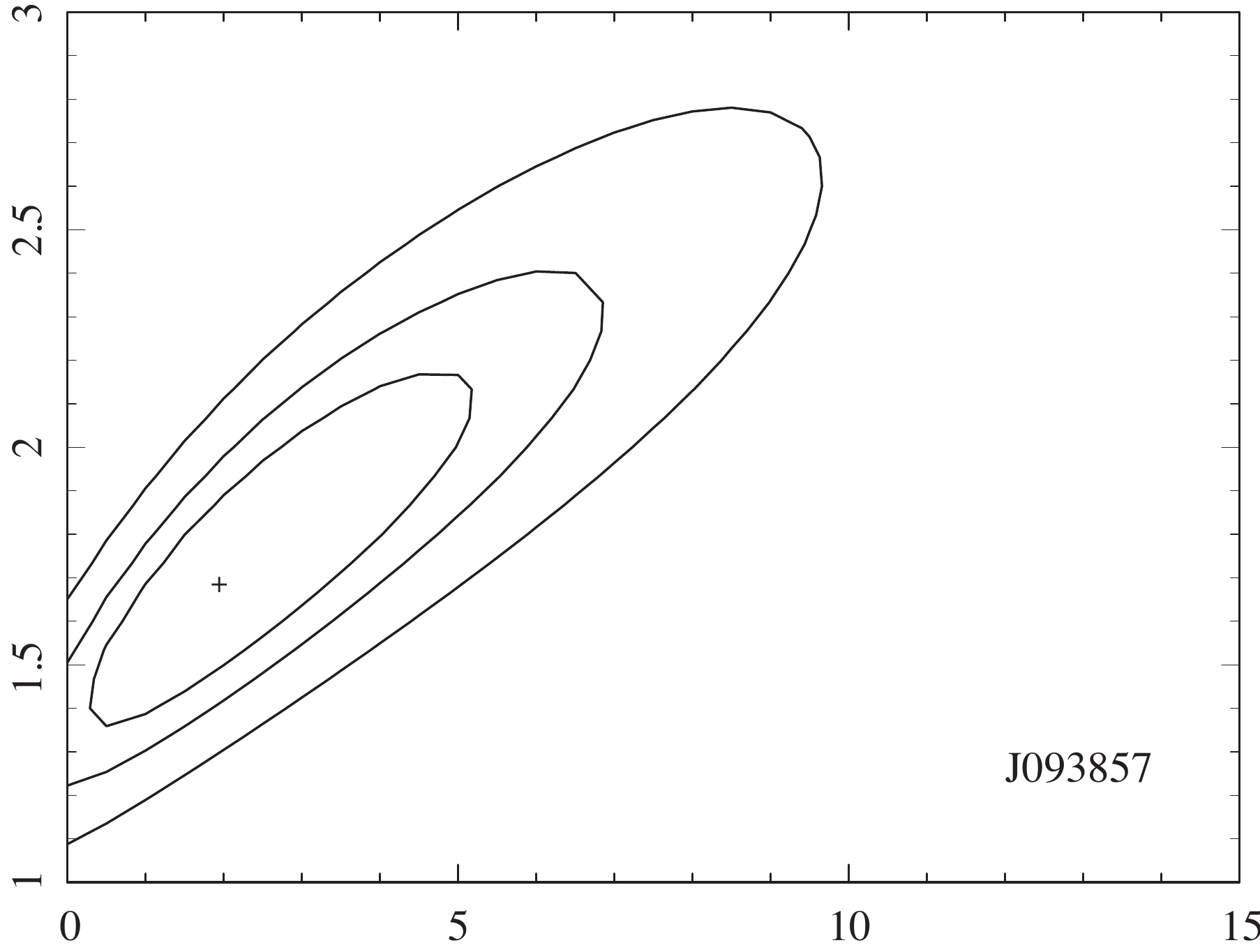}
 \includegraphics[scale=0.31,angle=0.0]{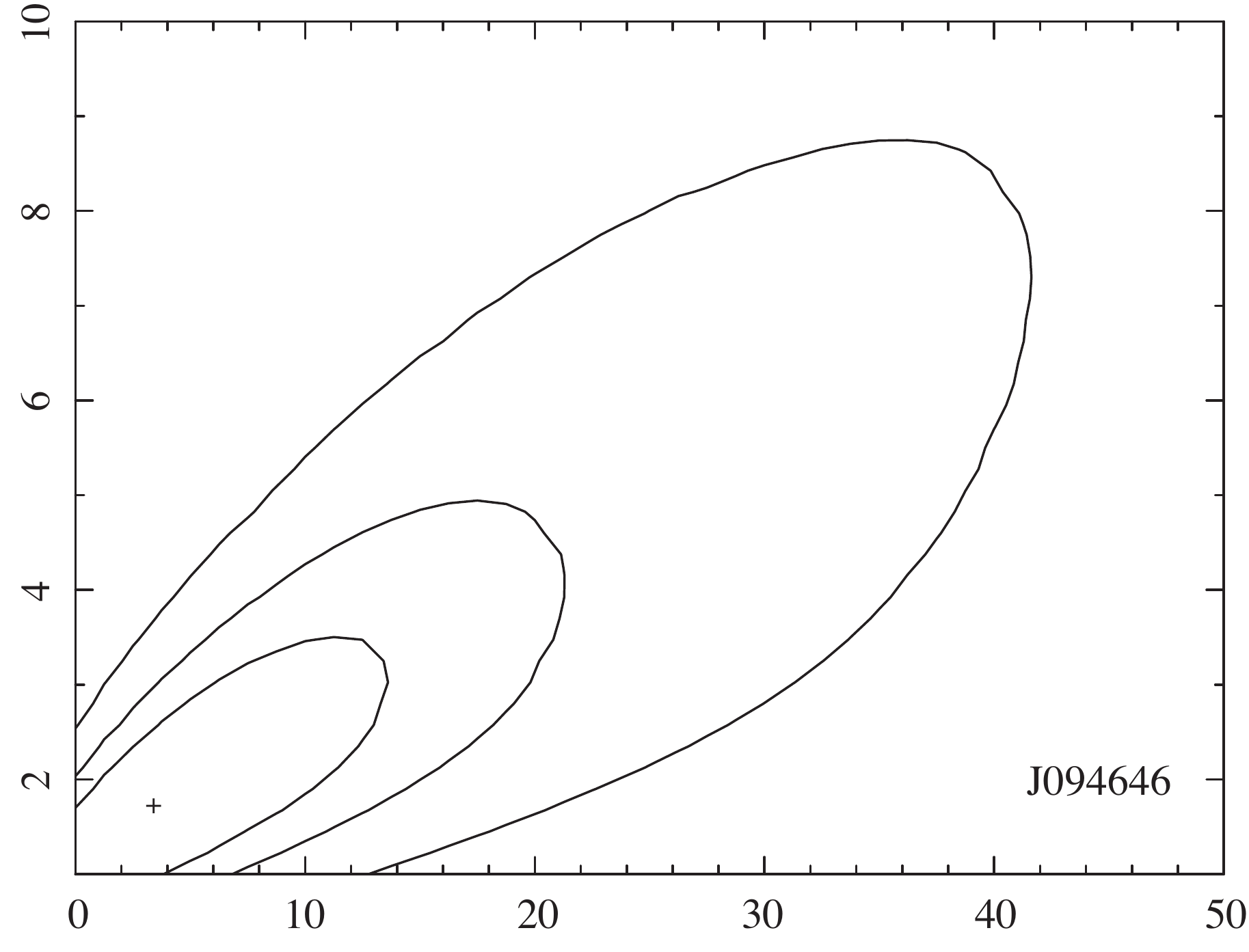}
 \includegraphics[scale=0.31,angle=0.0]{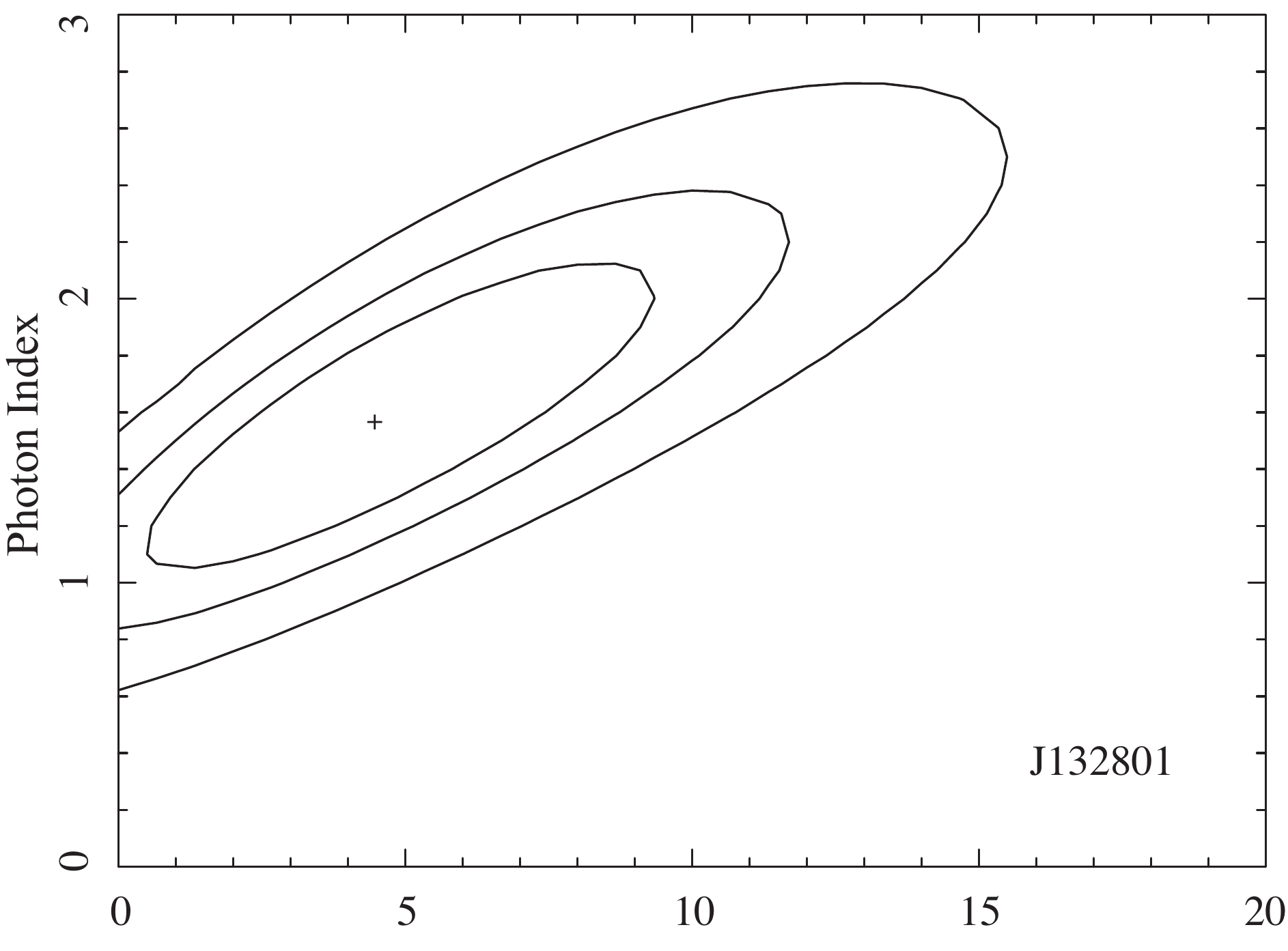}
 \includegraphics[scale=0.31,angle=0.0]{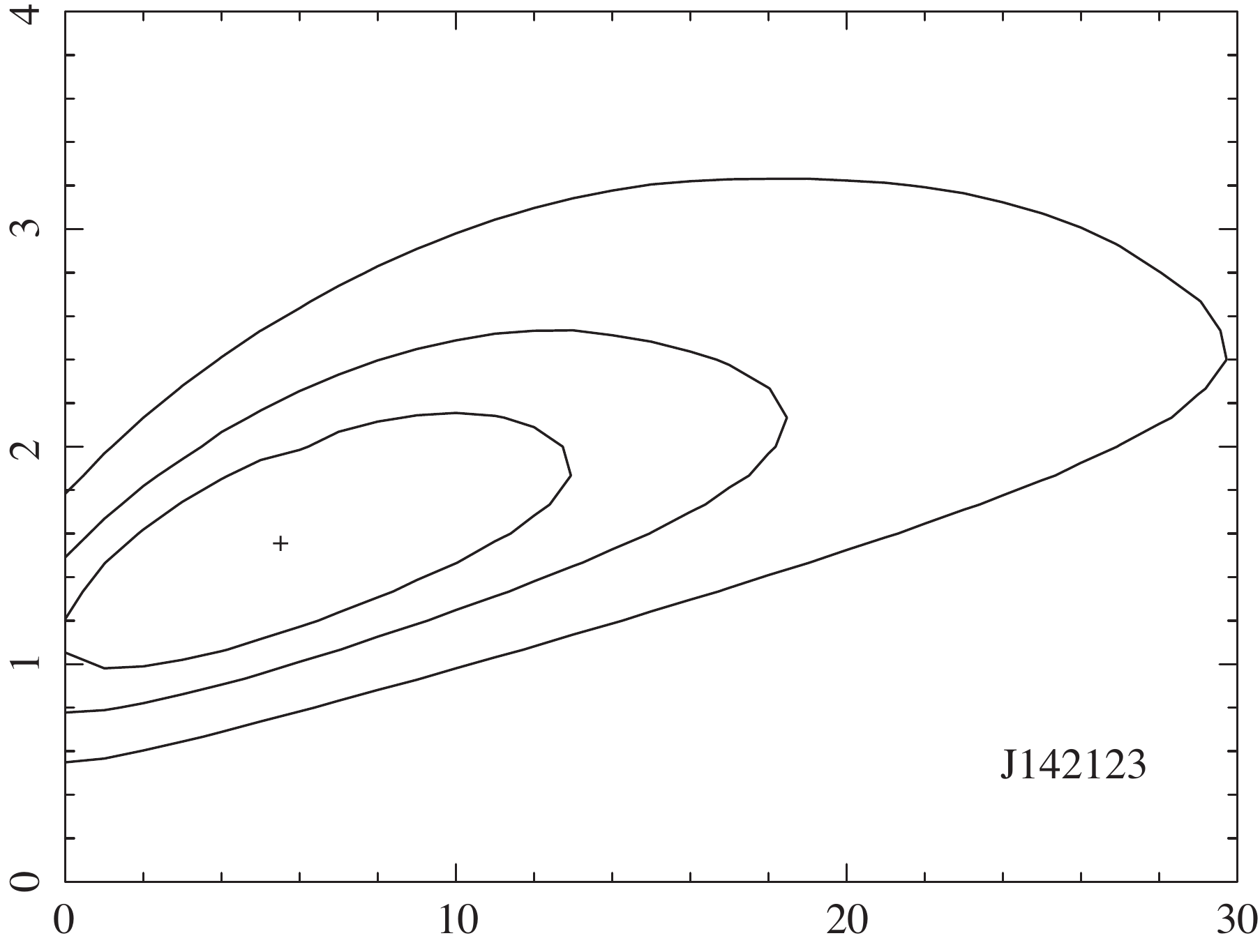}
 \includegraphics[scale=0.31,angle=0.0]{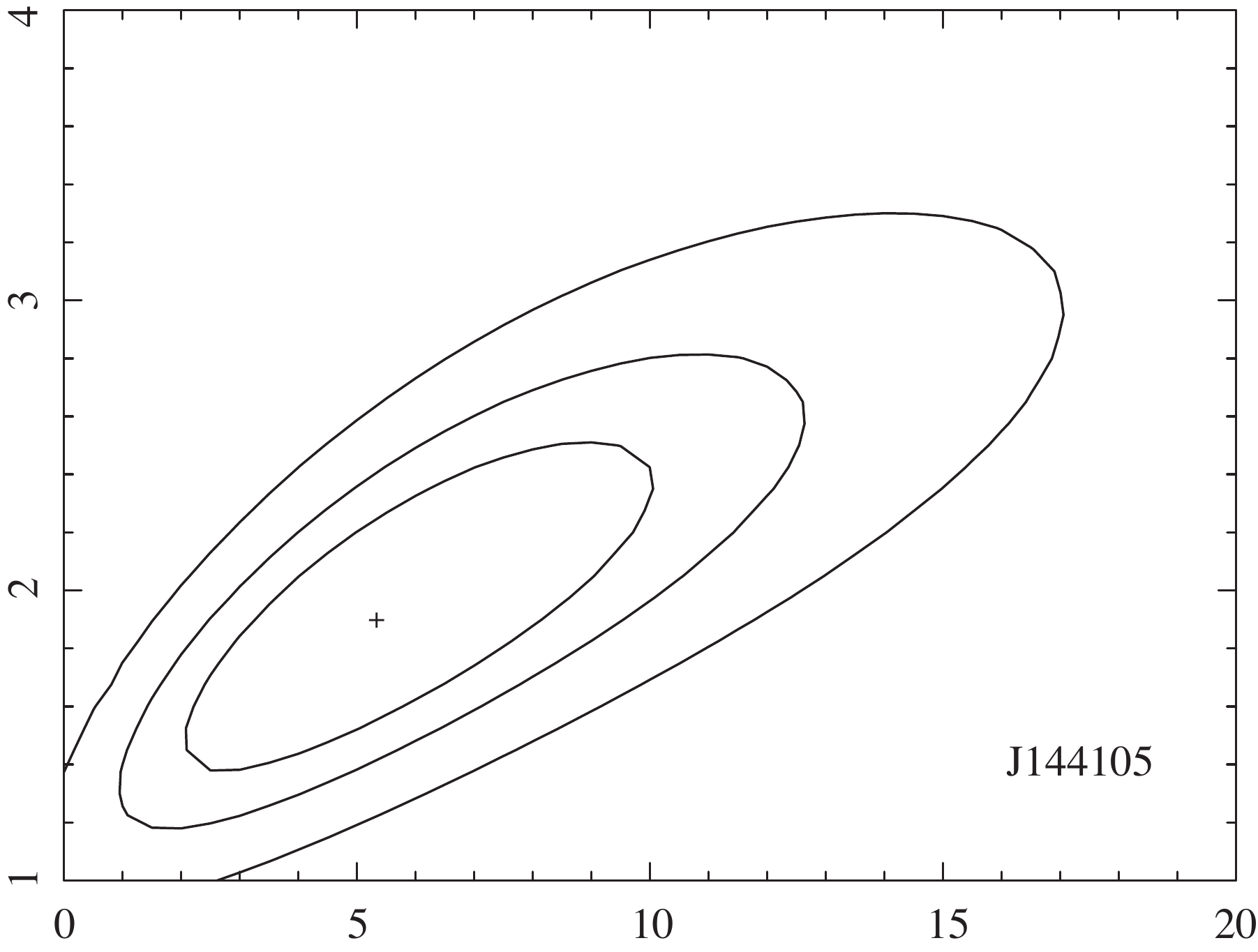}
\includegraphics[scale=0.31,angle=0.0]{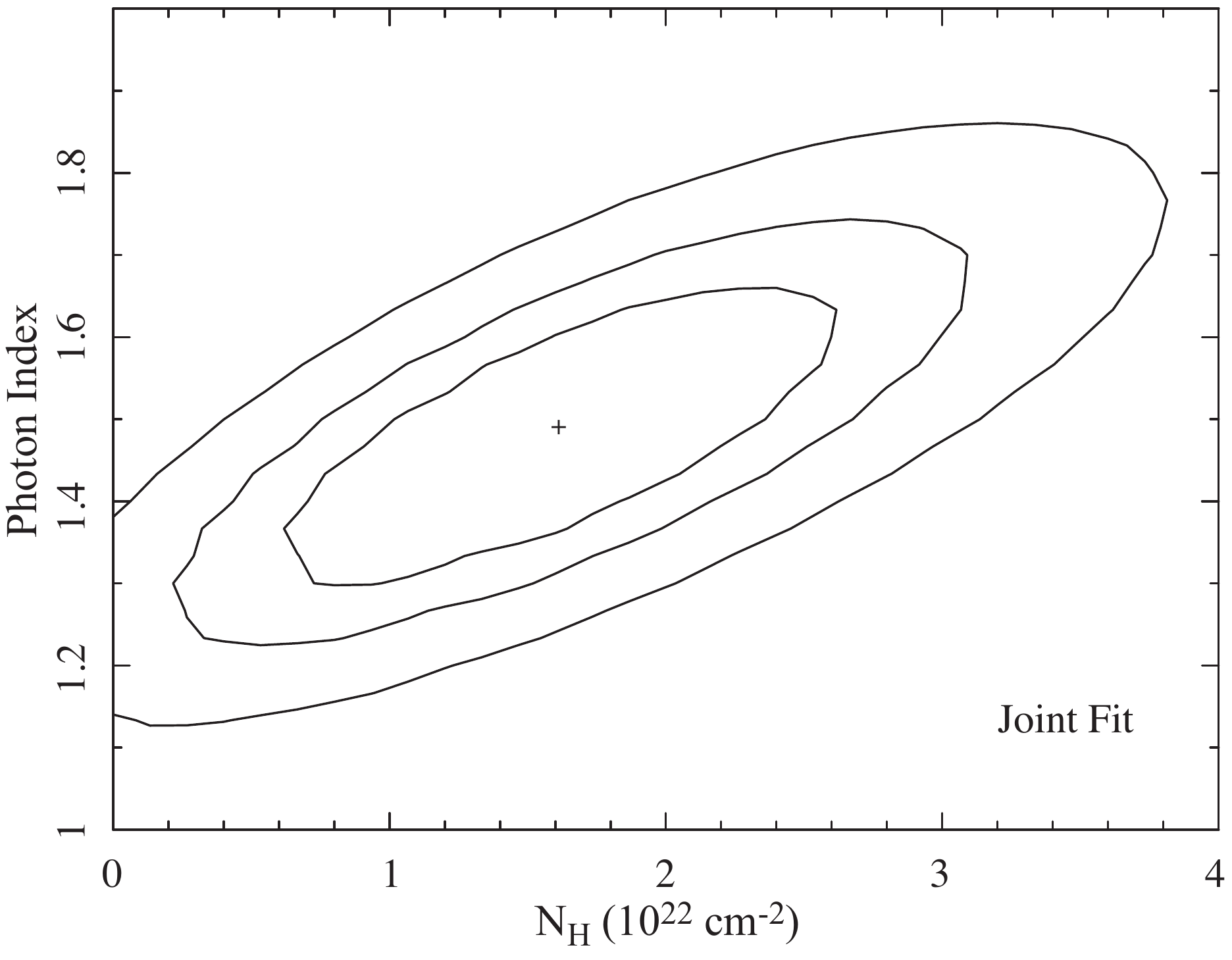}
\includegraphics[scale=0.31,angle=0.0]{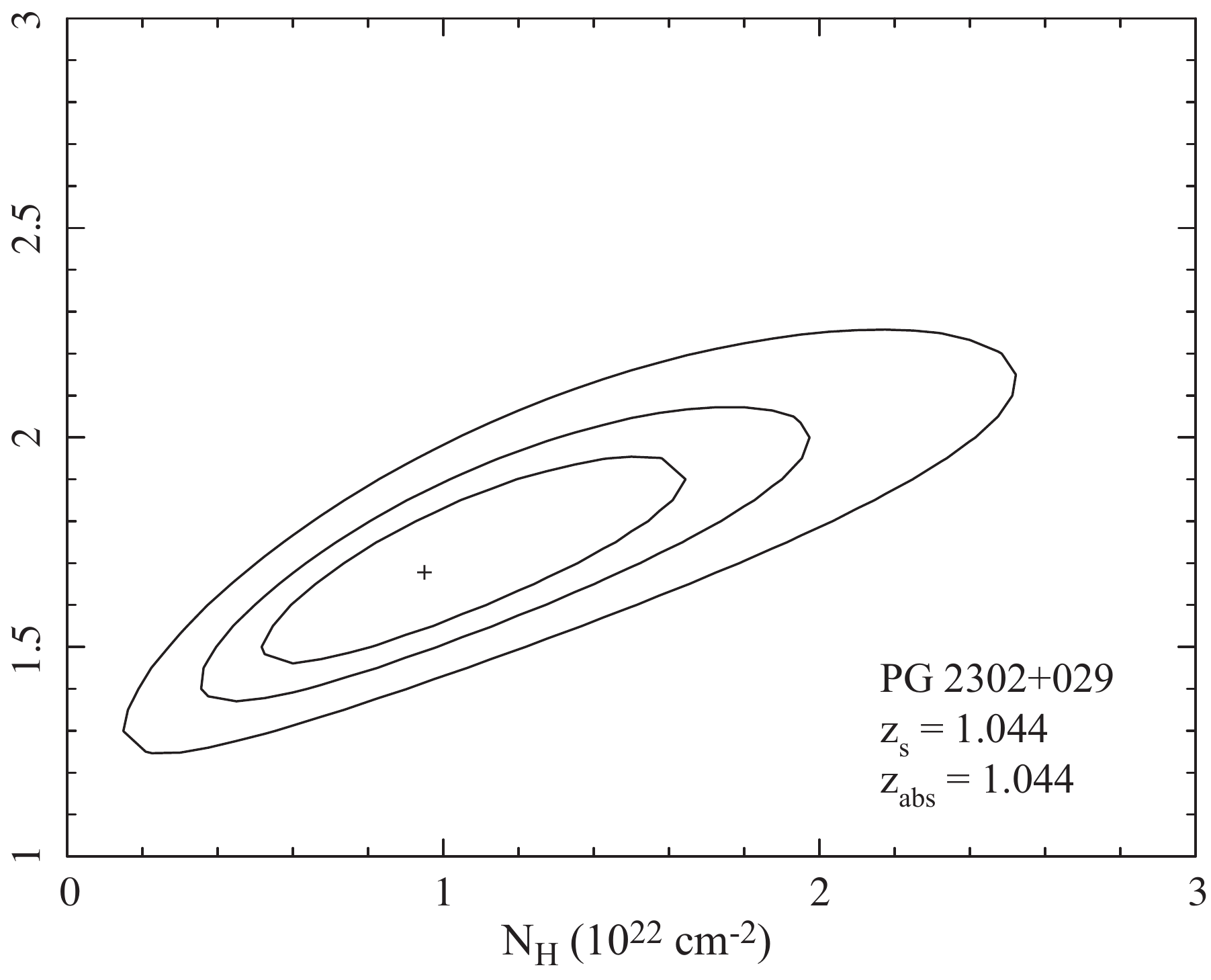}
\includegraphics[scale=0.31,angle=0.0]{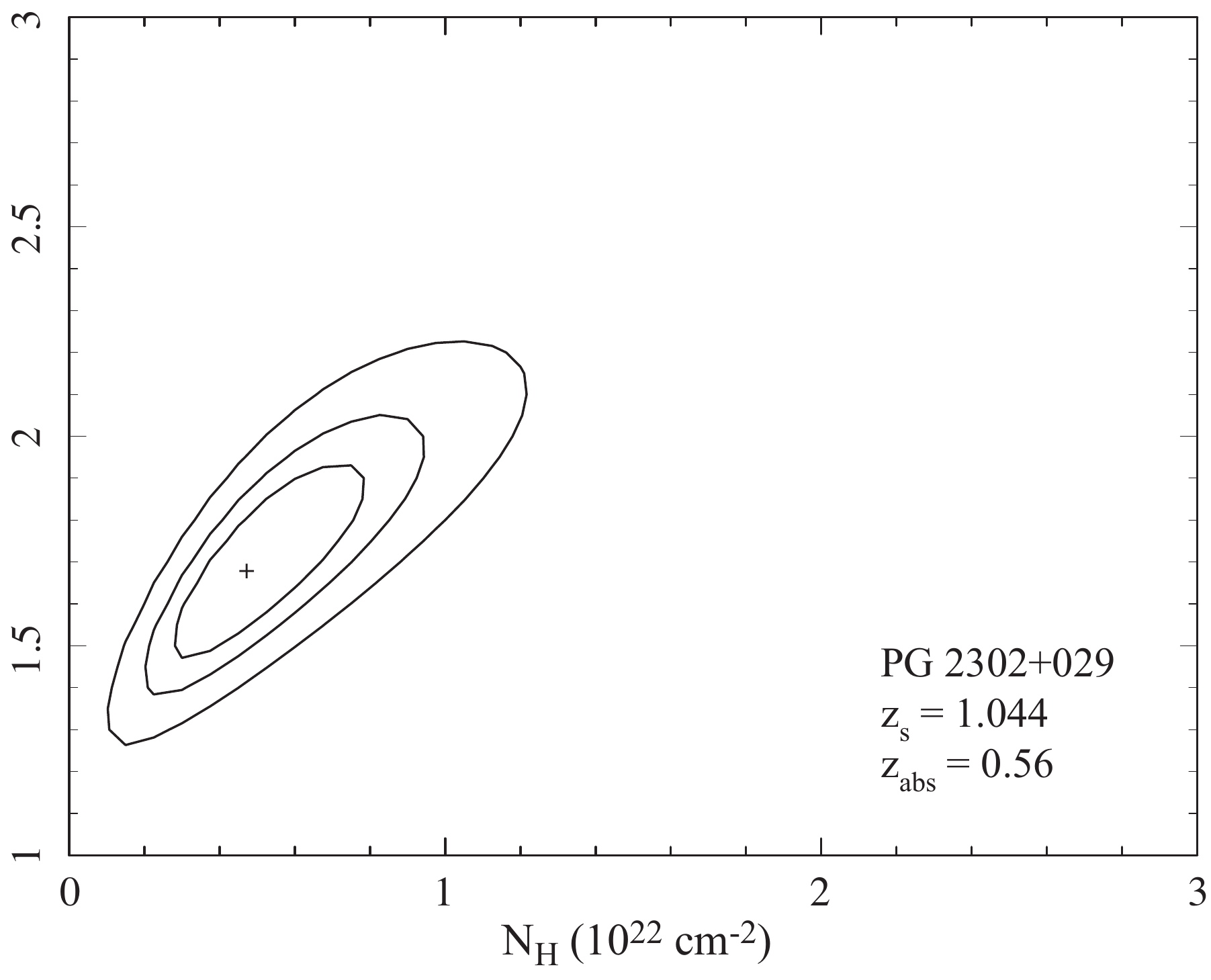}
 \end{center}
\vspace{-12pt}
 \caption{68\%, 90\%, and 99\% $\chi^2$ confidence contours in the X-ray photon index $\Gamma$ and total column density $N_H$ derived from spectral fits to the 7 quasars clearly detected (excluding J084255+331822), and from a joint analysis of 6 quasars from our new Chandra sample (excluding PG 2302+029 and the marginal mini-BAL J090508). The `+' in each panel marks the best fit values. The two plots for PG 2302+029 fix the absorber redshift at $z_{abs}=z_{em}$ (bottom middle panel) and at $z_{abs}=0.56$ favored by our $\chi^2$ analysis in Figure 3. }
\end{figure*}

The results for J093857+412821 in Figure 2 are consistent with our own previous measurement using the {\sl XMM Observatory} \citep{Paola11}. Our fit to those data indicate $\alpha_{ox} = -1.71$ and a neutral equivalent column density $\log(N_H/{\rm cm}^{-2})\le 22.3$ at 90\% confidence. That measurement is important because it was obtained in 2007.34, just 4 months after a ground-based spectrum showing the \civ\ mini-BAL still clearly present \citep[see][]{Paola11}. Thus the weakness of the X-ray absorption in the present study is not due to the mini-BAL fading away, as might be indicated by our MDM spectrum in  Figure 1 (see \S4.1 above). The X-ray absorption in this source was weak even when the mini-BAL was clearly present. 

We also perform a joint spectral fit to the 6 bona fide mini-BAL quasars in our sample with new {\sl Chandra} data, e.g., excluding PG 2302+029 and the marginal mini-BAL J090508+074151. The joint fit has the benefit of improving the signal-to-noise at the expense of averaging over redshifts and giving less weight to fainter sources. The results are shown in the lower left panel of Figure 2. Values of $\log(N_H/{\rm cm}^{-2}) \approx 22.2$ and $\Gamma\approx 1.5$ are preferred. \cite{Wu10} similarly performed a joint spectral fit to their core sample of 12 mini-BAL quasars measured with {\sl Chandra}. They derive $\Gamma\approx 1.9$ and $\log(N_H/{\rm cm}^{-2}) \la 21.9$ at 90\% confidence.

PG 2302+029 is a special case with extensive fitting already performed on {\sl Chandra} spectra by \cite{Sabra03}. They considered ionized absorbers at both the quasar redshift, $z_{em} \approx 1.044$, and the mini-BAL outflow redshift, $z_{mBAL} \approx 0.695$ corresponding to $v\sim 56,000$ \kms . They slightly preferred solutions with the X-ray absorber at $z_{mBAL}$ and total column density $\log (N_H/{\rm cm}^{-2})\approx 22.4$. Here we reanalyze those data using neutral absorbers for comparison to other results above. For the first time, we also allow the X-ray absorber redshift, $z_{abs}$, to be a free parameter. Figure 3 shows the $\chi^2$ statistic indicating goodness of fit as a function of $z_{abs}$ with $\Gamma$, $N_H$, and the flux normalization also allowed to vary. The figure shows significant $\chi^2$ minima near both the quasar redshift and the mini-BAL redshift, with redshifts near $z_{mBAL}$ slightly favored \citep[as in][]{Sabra03}. This raises the important question of whether the X-ray absorber is at the UV outflow velocity. The bottom right panels in Figure 2 show confidence contours for fits with the absorber redshift fixed at $z_{em}$ and $z_{abs} = 0.56$. In both cases, the total column densities are small, with $\log (N_H/{\rm cm}^{-2})\sim 22.0$ and $\sim$21.7, respectively. Note that these neutral-equivalent column densities are, as expected, about 0.7 dex smaller than the columns derived by \cite{Sabra03} for ionized absorbers.

\begin{figure}
\begin{center}
 \includegraphics[scale=0.35,angle=0.0]{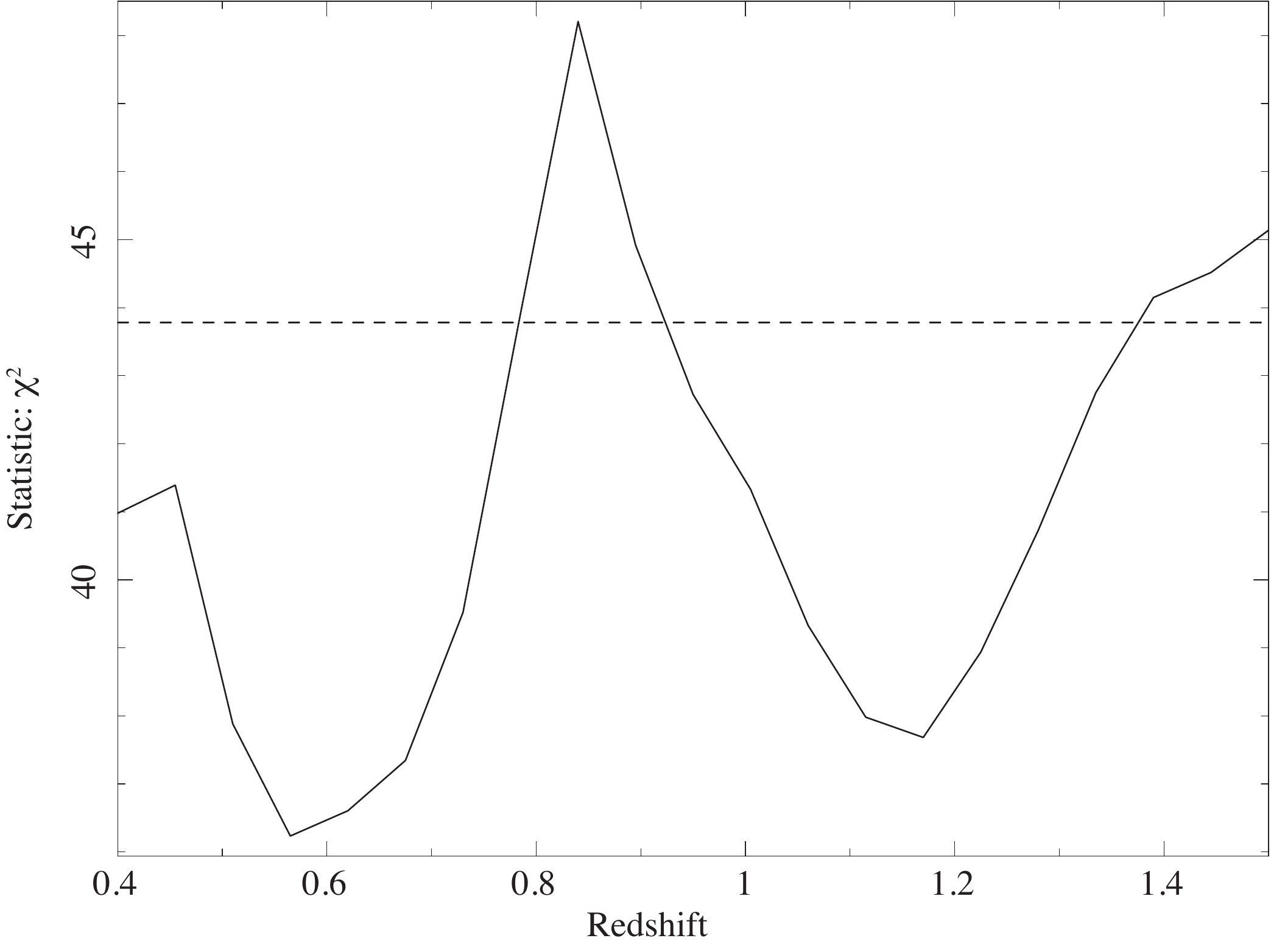}
 \end{center}
\vspace{-12pt}
 \caption{$\chi^2$ statistic versus X-ray absorber redshift, $z_{abs}$, from fits to the X-ray spectrum in PG 2302+029. There are significant minima near (but not at) the quasar redshift, $z_{em} \approx 1.044$ and the UV outflow redshift, $z_{mBAL}\approx 0.695$, with $z_{abs}\sim 0.6$ slightly preferred.}
\end{figure}

Spectral fits to BAL quasars in the literature indicate typically stronger X-ray absorption than mini-BALs, with $\log(N_H/{\rm cm}^{-2}) \ga 22.8$ \citep[e.g.,][]{Mathur00,Chartas02,Gallagher02,Gallagher06}. It is important to note that BALs exhibit a range of absorption properties \citep[see also][]{Fan09} and the comparisons between BAL and mini-BAL quasars are not straightforward. Part of the problem is ambiguous/different distinctions between BALs and mini-BALs in different studies (\S2). There are also uncertainties in the degree of ionization and line of sight covering factors that can contribute to a spread in derived $N_H$ assuming neutral absorbers \citep{Gallagher06,Gallagher07}. BAL quasars are also often undetected in X-rays, and therefore published fit data can be biased toward less absorbed sources with generally weaker UV BALs \citep[see also][]{Grupe03,Fan09}. \cite{Green01} derive the particular value $\log(N_H/{\rm cm}^{-2}) \sim 22.8$ from a joint fit to 6 high-ionization BAL quasars. Low-ionization BALs have consistently much stronger absorption with $\log(N_H/{\rm cm}^{-2}) \ga 24$ \citep[see also][]{Gallagher06,Morabito11}. 

Altogether, the spectral fits corroborate the results from $\Delta\alpha_{ox}$ and HR above: Extreme-velocity mini-BAL quasars have weak or moderate X-ray absorption, consistent with mini-BALs at lower speeds but weaker than BAL quasars by typical factors of at least several in $N_H$. These result are also consistent with NAL outflow studies, which indicate even less X-ray absorption than mini-BALs across a similar wide range of outflow speeds \citep{Misawa08,Chartas09,Chartas12,Hamann11}.

\section{Cloudy Simulations \& Limits on Radiative Shielding}

In this section, we perform photoionization simulations to constrain the amount of radiative shielding by more realistic ionized absorbers. Our goal is to test whether the shielding consistent with our data is sufficient to regulate the ionization of the mini-BAL gas and facilitate its radiative acceleration (\S1). We assume the shielding gas is located at the base of the outflow, between the continuum source and the mini-BAL outflow. We also assume the shield is approximately at rest (or at least not part of the extreme-speed mini-BAL flows) because it is too ionized and too transparent to be radiatively driven to high speeds. This assumption is supported by explicit calculations of the opacities and radiative forces \citep[][also Hamann et al., in prep.]{Murray95}. It is also a core premise of the models we are trying to test. (A high-column density shield moving at extreme speeds like the mini-BALs would undermine the premise of radiative driving because this gas would be highly ionized, nearly transparent, and some other mechanism would be needed to accelerate it; see also \S6 below.) With these assumptions, the amount of shielding across the UV to X-ray spectrum is tightly constrained by the X-ray data and by the non-detections of near-UV absorption lines near zero velocity that might form in the shield\footnote{Two of the quasars in our sample, J093857+412821 and J094646+392719, have narrow ``associated" absorption lines (AALs) within a few thousand \kms\ of the quasar redshifts. The origin of the these lines is not known, but in large quasar samples the presence of AALs does not correlate with broad outflow lines at higher speeds. We assume they are unrelated to the mini-BAL outflows.}. 

We use the photoionization and spectral synthesis code Cloudy \citep[version 10.00, last described by][]{Ferland98} to simulate spectra transmitted through hypothetical shielding slabs. The incident spectrum is shown by the solid black curves in Figure 4, scaled to match the luminosity $\nu\,L_{\nu}(2500{\rm A}) = 8\times 10^{46}$ ergs s$^{-1}$ typical of quasars in our sample. Its main features are powerlaw slopes across optical-UV and X-ray wavelengths of $\alpha_{uv}=-0.5$ and $\alpha_x=-0.9$, respectively, for $f_{\nu}\propto\nu^{\alpha}$. These powerlaw segments are joined smoothly in the far-UV by an exponential Wien function with temperature $T=200,000$ K. The relative strengths of the UV and X-ray spectral segments are scaled to yield $\alpha_{ox} = -1.7$, which is typical of radio-quiet quasars with 2500 \AA\ luminosities like our sample \citep[Table 1,][]{Steffen06}. Overall, this spectral shape\footnote{The Cloudy command used to define this continuum is: {\tt AGN T=200000K, a(ox)= -1.7, a(uv)= -0.5, a(x)= -0.9}} is in good agreement with quasar observations \citep[e.g.,][]{Reeves00,Richards06,Hopkins07,Shull12}. However, our results are not sensitive to the continuum shape details. The spectrum cuts off sharply at low ($h\nu <0.1$ eV) and high ($h\nu > 100$ keV) energies not important for our calculations. Some useful bolometric correction factors for this spectral shape are $L = 4.0\,\nu L_{\nu}(2500{\rm A})$, $L = 3.7\,\nu L_{\nu}(1500{\rm A})$, and $L = 260\,\nu L_{\nu}(2{\rm keV})$.

The Cloudy simulations consider plane parallel absorbing regions that completely cover the background light source. (We discuss the effects of more complex geometries at the end of this section.) The calculations also assume solar metallicity and a constant gas density $n_H = 10^8$ \cmn , although the particular density is not important for our calculations \citep{Hamann97d,Leighly11}. The metallicity matters because the metals can dominate the far-UV continuum opacities. However, apart from metal poor regimes not expected for quasar environments \citep{Hamann99}, our shielding predictions can be applied to a wide range of metallicities by scaling the model $N_H$ values inversely with the metallicity to achieve the same metal opacities.

\subsection{Model Shields Constrained by Observations}

Table 3 and Figure 4 show results for four Cloudy simulations that make different assumptions about the absorber properties. The first two models, named `noC4b100' and `noC4b1000', maximize the amount of shielding while {\it not} allowing the shield to produce a significant \civ\ absorption line in the near-UV. This is achieved by adopting a large total column density, $\log(N_H/{\rm cm}^{-2}) = 23.5$, and then increasing the ionization parameter, $U$ (defined as the dimensionless ratio of the H-ionizing photon density to H particle density at the illuminated face of the clouds), until the stronger line in the \civ\ doublet, 1548 \AA , has line center optical depth $\tau(\civ) \approx 0.1$. This sets a reasonable upper limit on the strength of \civ\ lines that could be detected in our data. The relationship between line optical depth and ionic column density depends on the velocity field inside the absorbing medium. For simplicity, we assume random turbulence. The two models noC4b100 and noC4b1000 differ only in the value of the doppler parameter used to define the velocity spread, e.g., $b=100$ \kms\ for noC4b100 and $b=1000$ \kms\ for noC4b1000. 

\begin{table*}
 \centering
 \begin{minipage}{158mm}
  \caption{Theoretical X-Ray Absorber/Shielding Results}
  \begin{tabular}{@{}lccccccccccccccc@{}}
  \hline
 & & & & & & & & \multispan{4}{\hfil --------- HR ---------\hfil}\\
  & & & & \multispan{4}{\hfil Line Center Optical Depths\hfil}&  \multispan{2}{\hfil z = 2.0\hfil}& \multispan{2}{\hfil z = 3.3\hfil}& \multispan{2}{\hfil --- $\Delta \alpha_{ox}$ ---\hfil}\\ 
 ~~Name & $\log N_H$ & $\log U$ & $b$ & \civ & \nv & \ovi & \neviii & -0.9& -0.5& -0.9& -0.5& -0.9& -0.5\\
\hline
{\it Cloudy Models:}\\
~~noC4b100 & 23.5 & 2.11 & 100 & 0.1 & 2.5 & 226 & 365 & 4.8 & 6.3& 2.9 & 4.4& -0.64& -0.47\\
~~noC4b1000 & 23.5 & 2.09 & 1000 & 0.1 & 0.9 & 36 & 56 & 5.1 & 7.0& 3.0 & 4.6& -0.71& -0.56\\
~~noO6b1000 & 23.5 & 2.26 & 1000 & ... & ... & 0.1 & 3.5 & 3.0 & 4.2& 2.4 & 3.7& -0.43& -0.30\\
~~noC4b1000xr & 23.0 & 1.71 & 1000 & 0.1 & 1.0 & 39 & 65 & 2.7 & 4.8& 1.9 & 3.4& -0.32& -0.35\\
{\it Neutral Absorbers:}\\
~~neutral23 & 23.0 & ... & ... & ...& ...& ...& ...& 4.0 & 6.6& 2.5 & 4.2& -0.68& -0.68\\
~~neutral22.5 & 22.5 & ... & ... & ...& ...& ...& ...& 2.3 & 3.9& 1.6 & 2.9& -0.23& -0.23\\
~~neutral22 & 22.0 & ... & ... & ...& ...& ...& ...& 1.5 & 2.8& 1.2 & 2.3& -0.07& -0.07\\
\hline
\end{tabular}
$N_H$ and $b$ have units \cmN\ and \kms , respectively. The line center optical depths apply to the short wavelength components of the doublets \civ\ \lam 1548, \nv\ \lam 1239, \ovi\ \lam 1032, and \neviii\ \lam 770 with contributions from the other doublet component if there is blending at large $b$. Hardness ratios, HR, are listed for redshifts $z=2.0$ and $z=3.3$. HR and $\Delta\alpha_{ox}$ are given in pairs for incident spectra with $\alpha_x=-0.9$ and $\alpha_x = -0.5$. For comparison, unabsorbed spectra with $\alpha_x=-0.9$ and $\alpha_x = -0.5$ have HR = 0.85 and 1.81, respectively. 
\end{minipage}
\end{table*}

 \begin{figure*}
 \includegraphics[scale=0.395,angle=-90.0]{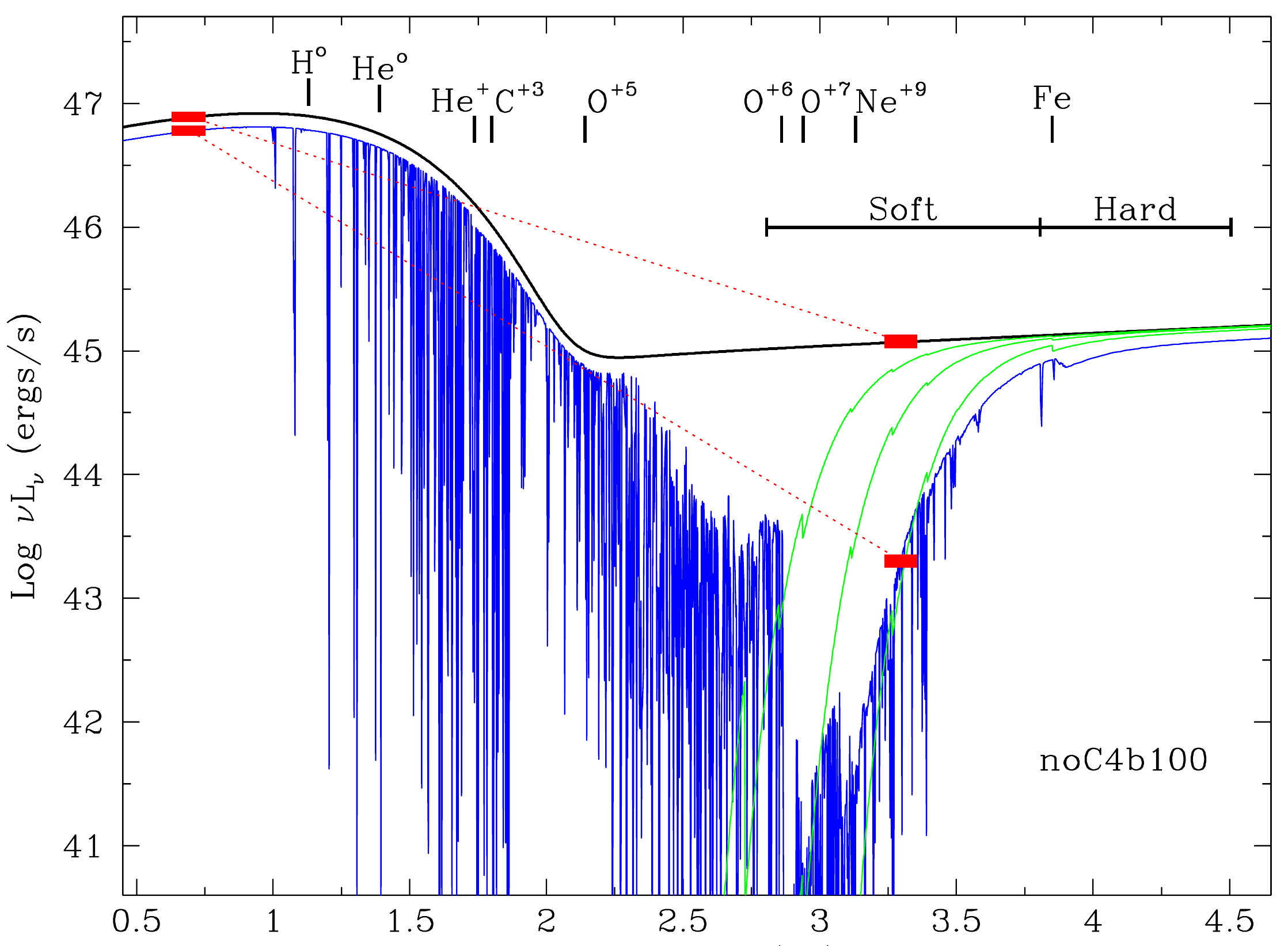}
 \includegraphics[scale=0.395,angle=-90.0]{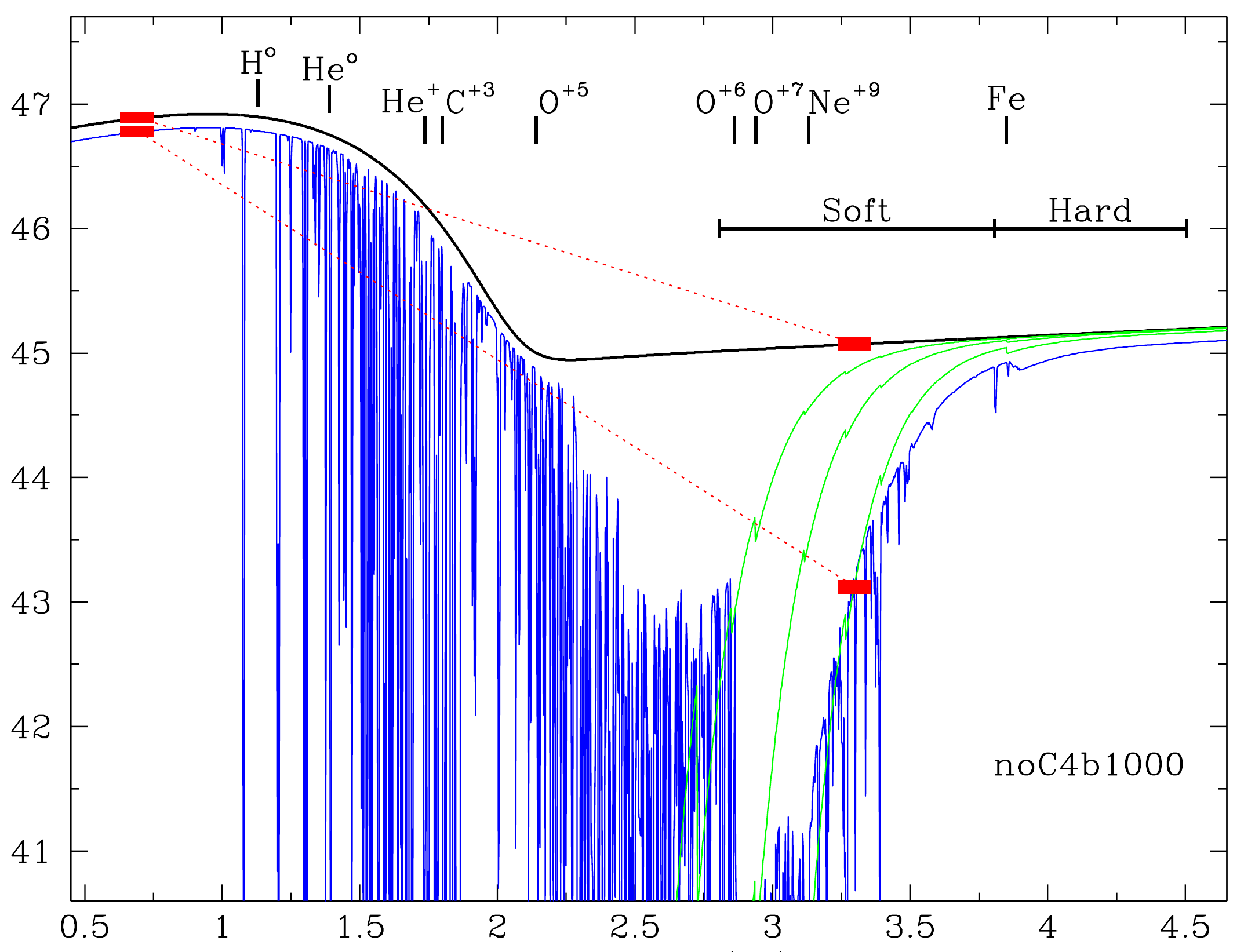}
 \includegraphics[scale=0.395,angle=-90.0]{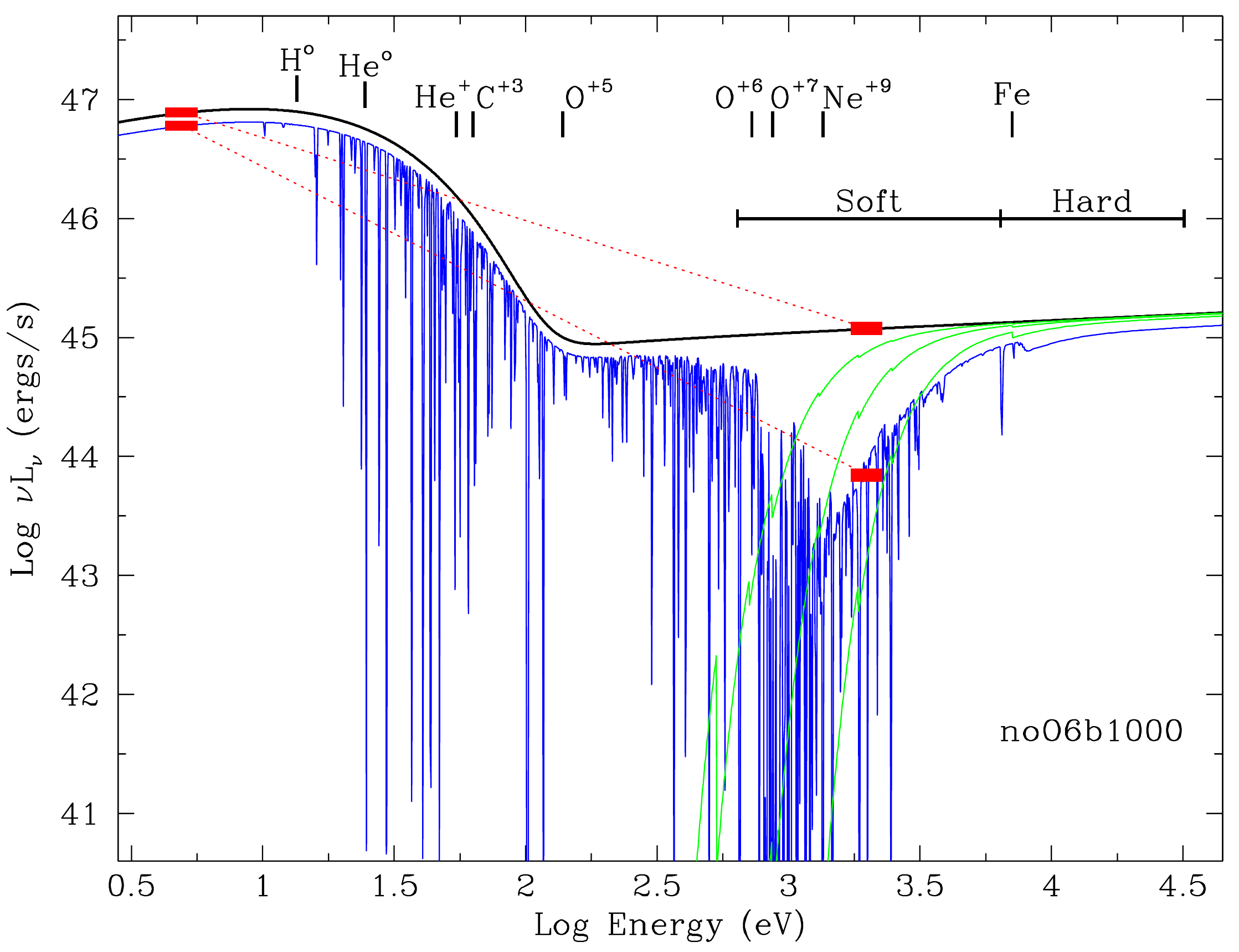}
\includegraphics[scale=0.395,angle=-90.0]{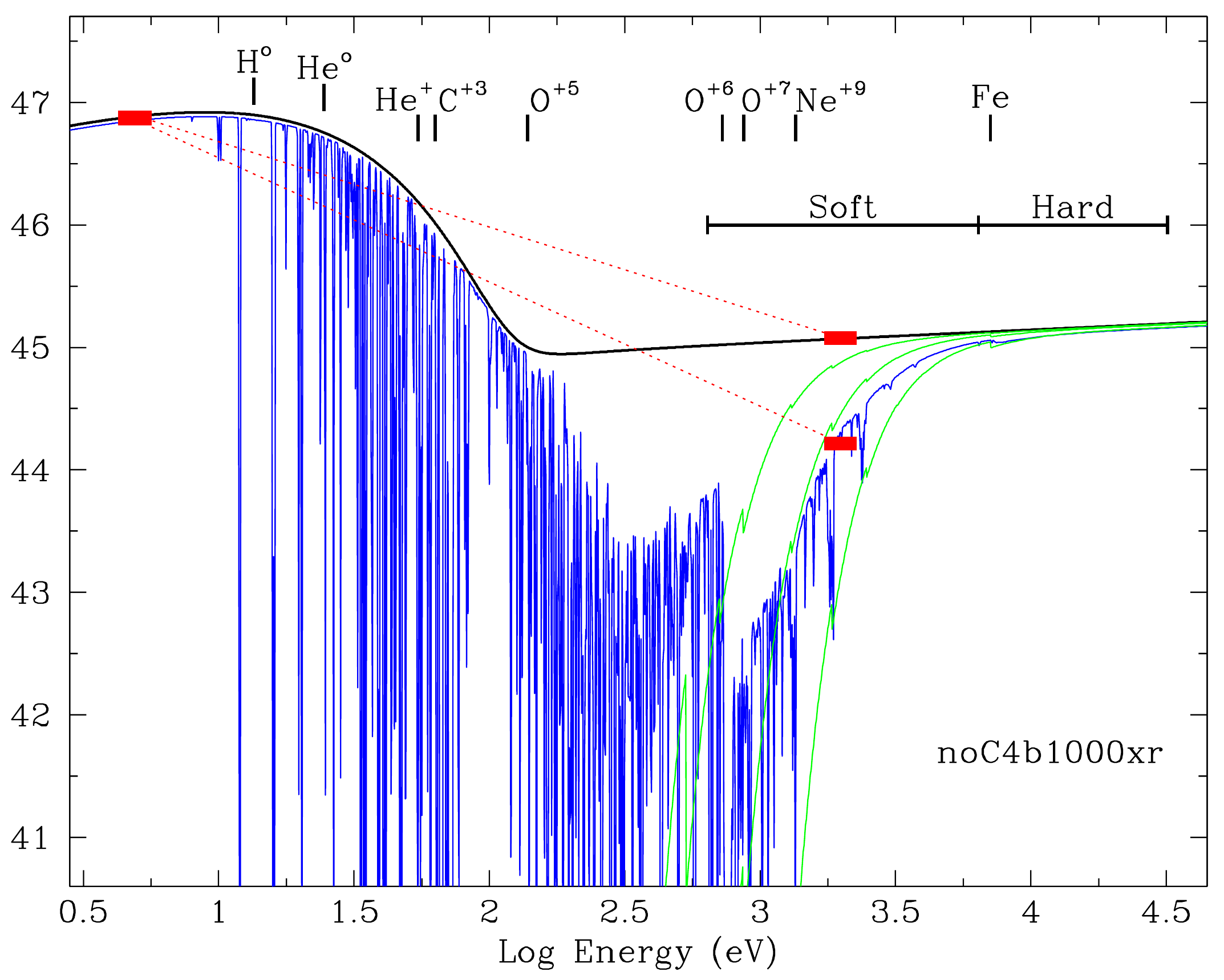}
\vspace{-3pt}
 \caption{Incident spectra (black curves) and transmitted spectra (blue curves) for the four Cloudy models with $\alpha_x = -0.9$ listed in Table 3. The model names are given in the lower right of each panel. Bold red dashes connected by thin red dotted lines show the luminosities at 2500 \AA\ (4.96 eV) and 2 keV used to measure $\alpha_{ox}$. The green curves in all panels show transmitted X-ray spectra for the neutral absorbers listed in Table 3, with $\log(N_H/{\rm cm}^{-2}) = 22$, 22.5, and 23 from left to right. Ionization energies for some important ions are marked across the top. The observed Soft (0.2-2.0 keV) and Hard (2.0-10 keV) X-ray bands are shown by horizontal dashes at redshift $z=2.2$ for illustration. Small offsets of the transmitted spectra below the incident spectrum are due to electron scattering. The spectra are sampled in increments of 1500 \kms\ and can therefore underestimate the line depths. See Table 3 and \S5.
   }
\end{figure*}

The top two panels in Figure 4 show the transmitted spectra for models noC4b100 and noC4b1000 (blue curves) compared to the incident spectrum (black). Strong absorption is evident across far-UV to soft X-ray wavelengths. This is caused by bound-free opacities plus many blended metal lines. The absorption is stronger in model noC4b1000 because the broader lines i) directly absorb more of the continuum flux, and ii) permit a lower degree of ionization and thus larger continuous opacities under the same $\tau(\civ) \approx 0.1$ constraint.

Table 3 lists the predicted hardness ratios, $\Delta\alpha_{ox}$ (measured relative to the incident spectrum), and line-center optical depths for several important lines in the model spectra. These values of HR and especially $\Delta\alpha_{ox}$ are substantially larger (more negative for $\Delta\alpha_{ox}$) than the observations (Table 2 and \S4.2), indicating that the models produce much more X-ray absorption than actual mini-BAL outflows. Nonetheless, these model absorbers are not effective shields for \civ\ and \ovi\ because they do not significantly suppress the flux near the C$^{+3}$ and O$^{+5}$ ionization edges (Figure 4). 

The models also predict very strong absorption lines of \ovi\ and \neviii , with optical depths $\ga$36 and $\ga$56 respectively.  Measurements or upper limits on these high ionization lines in actual quasars would place even stronger constraints on the overall amounts of shielding. Model `noO6b1000' in Table 3 and Figure 4 illustrates this point for an absorber with the same total column density, $\log(N_H/{\rm cm}^{-2}) = 23.5$, but now with the \ovi\ \lam 1032 line center optical depth constrained to the value $\tau(\ovi) \approx 0.1$ for  $b=1000$ \kms . This limit on $\tau(\ovi)$ requires a higher degree of ionization compared to the models constrained by $\tau(\civ) \approx 0.1$, resulting in weaker absorption at all wavelengths and completely negligible shielding. Nonetheless, the X-ray absorption at $\sim$2 keV is still stronger (more negative $\Delta\alpha_{ox}$) than most of the observations (Table 2 and \S4.2).

Model `noC4b1000xr' roughly maximizes the far-UV shielding consistent with our data by requiring $\tau(\civ) \approx 0.1$ for $b=1000$ \kms\ (as in model noC4b1000 above) with an added constraint on the total column to yield $\Delta\alpha_{ox} = -0.32$, which is at the high end of measured values in mini-BAL quasars (\S4.2). The resulting spectrum shown in the lower right panel of Figure 4 shows that the strength of X-ray absorption is similar to a neutral absorber with $\log(N_H/{\rm cm}^{-2})\sim 22.7$ (see also $\Delta\alpha_{ox}$ in Table 3). This model again predicts a strong saturated \ovi\ and \neviii\ lines near the rest velocity, which violates previous observations of J093857+412821 \citep{Paola11} and some other well-measured cases of NAL and mini-BAL outflows \cite{Jannuzi96,Sabra03,Telfer98,Hamann11,Paola11}. Nonetheless, there are still only modest amounts of far-UV and soft X-ray absorption that are again not sufficient to shield the mini-BAL gas. 

\subsection{How the Model Shields Fail}

We can quantify the failure of the model absorbers in \S5.1 to be effective shields by examining their influence on the ionization of gas behind them. First, we consider hypothetical mini-BAL regions exposed directly to the quasar continuum described above, with no shielding at all. From the definition of the ionization parameter, 
\begin{equation}
U \ \equiv \ {{Q_H}\over{4\pi c\, R^2\, n_H}}
\end{equation}
where $Q_H$ is the total emitted luminosity of hydrogen-ionizing photons (\#/s), we have this general relationship between the gas density, $n_H$, and its radial distance, $R$, from the continuum source, 
\begin{equation}
n_H \; = \; 4\times10^8 \left({{\nu L_\nu (2500{\rm \AA })}
\over{8\times 10^{46}\,{\rm ergs/s}}}\right)
\left({{0.4}\over{U}}\right)\left({{2\,{\rm pc}}\over{R}}\right)^2
~{\rm cm}^{-3}
\end{equation}
where the luminosity $\nu L_\nu (2500{\rm \AA }) = 8\times 10^{46}$ ergs s$^{-1}$ is roughly typical of our sample and $R = 2$ pc is a reasonable guess for the location of the mini-BAL gas (\S4.1). If the mini-BAL gas is optically thin throughout the Lyman continuum, this equation and the continuum shape fully describe its ionization. The value of $U\approx 0.4$ is conservatively high for \civ\ and \ovi\ absorption, slightly favoring \ovi\ to be consistent with mini-BAL observations (see refs. above). The specific ion fractions are C$^{+3}$/C $\approx  0.06$, O$^{+5}$/O $\approx 0.16$, and C$^{+3}$/O$^{+5} \approx 0.26$ for solar abundances. We conclude from Equation 2 that high densities of order $n_H\sim 4\times 10^8$ \cmn\ are needed to keep the ionization low enough for \civ\ absorption if a shield is not present.

Figure 5 shows similar results for more realistic situations with non-negligible column densities in the mini-BAL region. In particular, the solid curves show the values of $U$ and $n_H$ that produce \civ\ (black) or \ovi\ (blue) mini-BALs with $\tau\approx 0.1$ and $b=1000$ \kms\ for different total column densities, $N_H$. These curves thus define the minimum densities (and maximum $U$) needed for significant/measurable \civ\ and \ovi\ mini-BALs in outflow regions without a shield. The densities shown in this figure are for the particular distance $R=2$ pc, but they can be scaled to other distances by multiplying by the factor (2 pc/$R$)$^2$, as in Equation 2. 

\begin{figure}
\begin{center}
 \includegraphics[scale=0.45,angle=0.0]{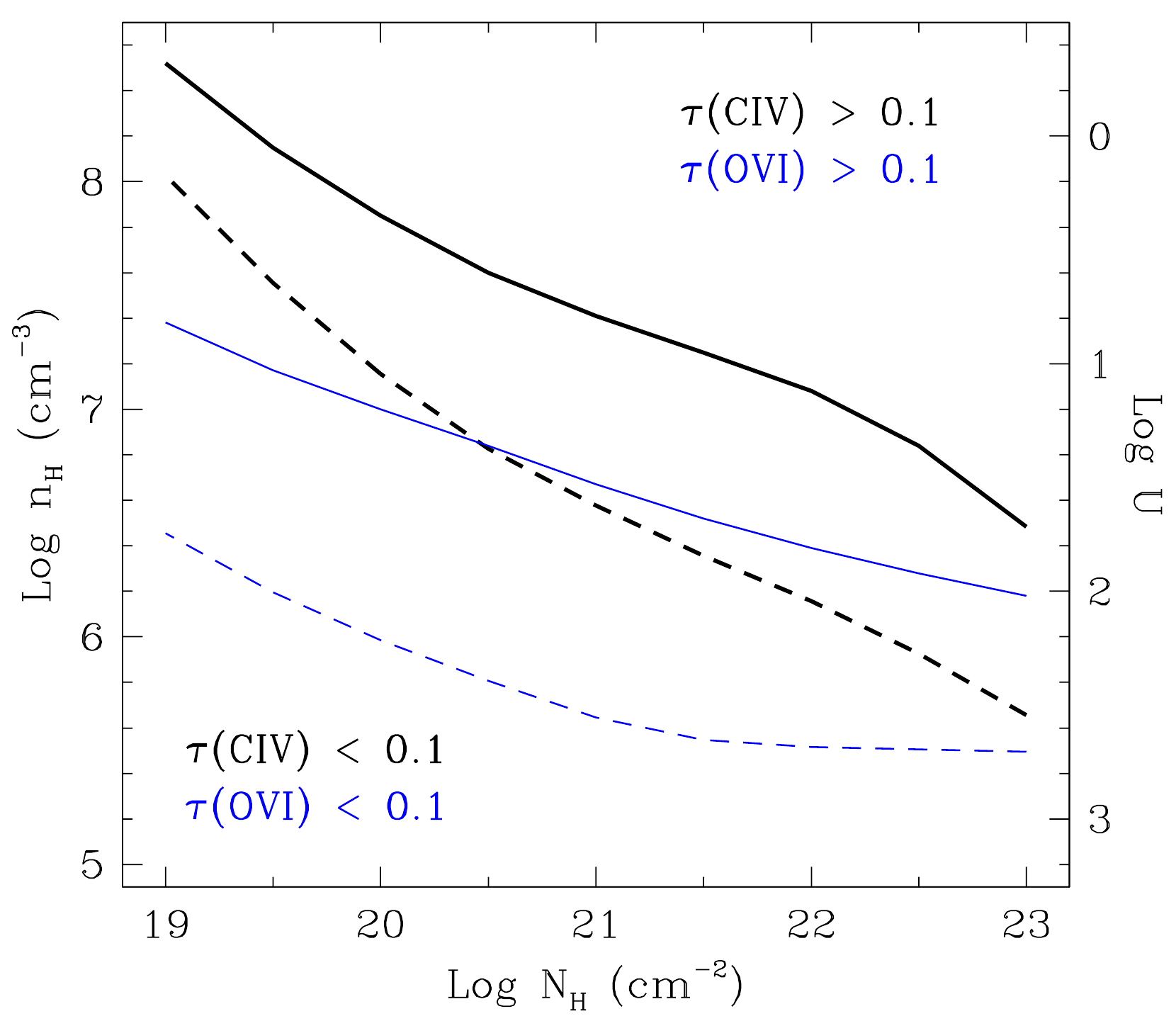}
 \end{center}
\vspace{-12pt}
 \caption{Densities, $n_H$, and ionization parameters, $U$, needed to produce \civ\ or \ovi\ mini-BALs in outflow regions with different total column densities, $N_H$. The specific densities shown are for $R=2$ pc, but they can be scaled to any distance by the factor $(2\,{\rm pc}/R)^2$. The solid bold/black and thin/blue curves show the conditions needed for $\tau\approx 0.1$ mini-BALs of \civ /\ovi\ (for $b=1000$ \kms ) in outflow gas illuminated by the unattenuated quasar spectrum (Figure 4). Densities above these curves lead to stronger/measurable mini-BALs. The dashed bold/black and thin/blue curves mark $\tau\approx 0.1$ absorption by \civ /\ovi\ in mini-BAL regions behind a maximum defined by model noC4b1000xr (Table 3). The $U$ values appropriate for the dashed curves are $\sim$0.12 larger than shown on the right-hand axis (because $Q_H$ for the transmitted shield spectrum is slightly reduced, Eqn. 1).
   }
\end{figure}

Next we insert the strongest shield consistent with our data (model noC4b1000xr) between the quasar continuum source and the mini-BAL gas. We do this using the transmitted spectrum from model noC4b1000xr (lower right panel in Fig. 4) to illuminate the hypothetical mini-BAL clouds described above. In the optically thin case (Eqn. 2), the ionization in the mini-BAL gas decreases to yield C$^{+3}$/C $\approx  0.24$, O$^{+5}$/O $\approx 0.06$, and C$^{+3}$/O$^{+5} \approx 2.0$. Ionization conditions similar to the unshielded situation can now occur at a factor of $\sim$3 lower density compared to Equation 2. 

For the cases with significant column densities in the mini-BAL region, the dashed curves in Figure 5 show the revised values of $n_H$ and $U$ needed for just-measurable \civ\ and \ovi\ mini-BALs with $\tau \approx 0.1$ behind the maximum shield. These results are approximate because the putative shield is not offset in velocity from the mini-BAL gas. Nonetheless, the displacement between the dashed and solid curves in Figure 5 indicates that the addition of the {\it maximum} shield allowed by our data reduces the density thresholds by factors of $\sim$3 to $\sim$10. These changes are significant but not nearly large enough to affect the structure of the flow behind the shield or facilitate the acceleration of outflow gas that would otherwise be too ionized for radiative driving (see also \S6 below). 

All of the models just described, including noC4b100 and noC4b1000 which grossly over-predict the X-ray absorption, fail to be effective shields because there is not sufficient continuous opacity near the ionization energies of C$^{+3}$ and O$^{+5}$ (Figure 4). Notably absent are significant absorption edges of H$^o$ (13.6 eV), He$^o$ (24.6 eV), or He$^+$ (54.4 eV) because the gas is too ionized. Other candidates for suppressing the flux near the C$^{+3}$ and O$^{+5}$ edges are the C$^{+3}$ and O$^{+5}$ ions themselves. These edges are also missing because the shields are not allowed to form \civ\ or \ovi\ absorption lines. This is evident from the relationship between $\tau(\civ )$ and the optical depth at the C$^{+3}$ edge, $\tau({\rm C}^{+3})$, namely,
\begin{equation}
\tau(\civ ) \ = \ 650\, \left({{b}\over{1000\,{\rm km/s}}}\right)\tau({\rm C}^{+3})
\end{equation}
and similarly for $\tau(\ovi )$ related to $\tau({\rm O}^{+5})$ at the O$^{+5}$ edge
\begin{equation}
\tau(\ovi ) \ = \ 570\, \left({{b}\over{1000\,{\rm km/s}}}\right)\tau({\rm O}^{+5})
\end{equation}
using atomic data from \cite{Verner96} and \cite{Osterbrock89}. Weak \civ\ and especially weak \ovi\ lines in the near-UV lead to very weak absorption at critical far-UV energies below $\sim$200 eV. Conversely, an effective shield that {\it does} have strong absorption at these energies will produce very strong \civ\ and/or \ovi\ absorption lines, which can be tested by UV observations. 

The absence of these lines near $v\sim 0$ points to the absence of a viable shield at $v\sim 0$. We cannot exclude the possibility that the mini-BAL regions themselves have large column densities leading to some amount of ``self-shielding.'' However, existing data do not support large mini-BAL or NAL outflow column densities\footnote{Very few estimates of $N_H$ are available for NAL and especially mini-BAL outflows in quasars. Our own analysis of J093857+412821 in the current sample \citep[][as measured in 1996]{Paola11} crudely indicates $N_H\sim 5\times 10^{19}$ \cmN\ in the mini-BAL gas, which is similar to other estimates for lower speed outflow lines in quasars \citep[e.g.,][]{Gabel06,Hamann11}. \cite{Wu10} find $N_H \la 10^{19}$ \cmN\ for some weaker and much lower speed NAL outflows. \cite{Telfer98} favored larger values, $N_H > 10^{21}$ \cmN , for one particular mini-BAL system that might have unusual properties.}. A self-shielding mini-BAL region would also imply large columns of highly ionized gas already moving at speeds $>$0.1c. This gas would be too transparent for radiative driving, which begs the question of how it was accelerated to high speeds without a separate shield (at $v\sim 0$, \S1). Moreover, within reasonable $N_H$ limits, a self-shielding mini-BAL region  would not change the main conclusion that high densities are needed to explain the observed ionizations low enough for \civ\ absorption lines (e.g., Figure 5, which considers mini-BAL column densities up to $N_H = 10^{23}$ \cmN ). We discuss the implications these density results in \S6 below. 

\subsection{Complex Geometries}

Here we address an important caveat: shielding gas might cover only portions of the spatially-stratified emitting regions. Some authors have argued that the strong X-ray absorption in BAL quasars is caused by a compact, high column density medium that covers the X-ray emission source (near the central SMBH) without covering the near-UV and visible continuum regions \citep[farther out in the accretion disk,][]{Gallagher07,Chartas09}. This picture is appealing for extreme cases like FeLoBALs where the X-ray absorber appears to be Compton thick even though the quasars are bright in the near-UV and visible \citep[][also Lou et al. 2013, ApJ, in press; Hamann et al., in prep.]{Rogerson11,Morabito11}. It might also explain sources where the outflow velocities measured in the UV and X-rays are dramatically different \citep[e.g.,][]{Chartas07,Giustini11}. 

The minimum requirement for a shield in any geometry is that it covers (at least) the far-UV emission source as seen from the outflow to prevent over-ionization. In BAL quasars, the observed high column density X-ray absorbers might reasonably cover the far-UV emission source without also covering the more extended near-UV and visible emission regions. This type of geometry might allow there to be significant shielding at rest in the quasar frame with no near-UV absorption lines near $v\sim 0$ (e.g., for BALs with detached troughs). However, in NAL and mini-BAL outflows the required geometry seems implausible because the shield would need to cover {\it only} the unobserved far-UV continuum source (essential for shielding) while not covering the near-UV and X-ray emitting regions (as seen by the observer). One can imagine geometries that might do this, but they all appear to have serious problems. A thorough discussion of these possibilities is beyond the scope of the present paper.

\section[]{Discussion}

The main result from \S5 is that NAL and mini-BAL outflows are not significantly shielded in the far-UV and, therefore, the outflow ionizations are not moderated by the presence of a shield. This disagrees with BAL outflow models that rely on shielding at the base of the outflow to maintain sufficient opacities for radiative driving (\S1). Evidently, NAL and mini-BAL outflows are accelerated to high speeds like BALs without the benefits of a radiative shield. 

One might suppose that this conclusion could be avoided by variable shielding, e.g., due to transverse motions of a patchy shielding medium near the continuum source \citep{Misawa10,Giustini11}. In that situation, the acceleration could occur during a past period of strong shielding, under BAL-like outflow conditions, while today we see the flow already at high speeds without a significant shield. However, this scenario does not explain the moderate degrees of ionization (\civ ) that we observe in the outflows now. As soon as the shield goes away, the resulting over-ionization of the outflow should appear (to an observer) roughly instantaneously because the absorption line gas lies along our lines of sight to the ionizing continuum source. 

Another possibility is that the NAL and mini-BAL gas was accelerated in a location with more shielding (e.g., nearer the accretion disk plane) before moving to its observed location (farther above the disk) where shielding is negligible \citep{Ganguly01}. This would be consistent with suggestions (based on variability) that mini-BALs form in small blobs or filaments along the ragged upper boundary of the main BAL outflow \citep{Hamann08,Hamann12}. However, it is not clear what vertical force would be available to push these mini-BAL blobs away from the disk plane {\it after} they have been accelerated to speeds $>$0.1c. Radiative forces can only drive the flow away from the UV continuum source, directly toward the observer. The flow trajectories might be curved by strong magnetic fields, but that describes a different model where magnetic forces also dominate the acceleration \citep{Fukumura10,Kazanas12}. 

The essential point is that NAL and mini-BAL outflows have moderate degrees of ionization, similar to BALs, even though they are not behind a radiative shield. If a shield is not needed to maintain these moderate outflow ionizations, then it is also not needed to facilitate the acceleration.  

The tendency for over-ionization can be seen by considering a smooth and continuous flow (volume filling factor unity) with radial thickness of order $\Delta R_{flow}\sim R$ at $R\sim 2$ pc. If this flow has a conservatively large total column density with $N_H\la 10^{22}$ \cmN , the volume density would be only $n_H\sim N_H/\Delta R_{flow}\la 1600$ \cmn . More realistic outflow columns with $N_H\la 10^{21}$ \cmN\ would imply $n_H\la 160$ \cmn . If there is no shielding, the ionization parameter in this gas would be conservatively $U > 10^5$ (ignoring minor effects of geometric dilution, Eqn. 2). Figure 5 shows that these flow densities would lead to complete over-ionization, with no detectable \civ\ or \ovi\ absorption, even if the flow resides behind the maximum shield allowed by our data. This gas would have no possibility for radiative acceleration to high speeds because it is too transparent \citep[][Hamann et al., in prep.]{Murray95}. 

We conclude that the outflow ionizations are kept moderate by high densities in discrete clouds or substructures. The required densities are shown in Figure 5. If the mini-BAL regions have conservatively $N_H \la 10^{22}$ \cmN\ at our fiducial radius $R\sim 2$ pc, then the minimum densities needed to produce a \civ\ mini-BAL are $n_H\ga 2\times 10^7$ \cmn\ if the gas is not shielded or $n_H\ga 2\times 10^6$ \cmn\ behind a maximum shield. The total radial extents of these mini-BAL clouds are only $\Delta R_{clouds} \approx N_H/n_H\la 5\times 10^{14}$ cm or $\Delta R_{clouds} \la 5\times 10^{15}$ cm in the unshielded and shielded cases, respectively. More realistic mini-BAL column densities $N_H \la 10^{21}$ \cmN\ require higher densities resulting in total cloud thicknesses $\Delta R_{clouds} \la 3\times 10^{13}$ cm or $\la$$3\times 10^{14}$ cm and radial filling factors of only $\Delta R_{clouds}/R \la 5\times 10^{-6}$ or $\la$$5\times 10^{-5}$ for the unshielded and maximum shield situations, respectively. 

These inferred cloud sizes and densities clearly depend on the outflow radius. At smaller radii, nearer the expected launch point of the outflows (\S4.1), the total cloud thicknesses and filling factors are much smaller because the minimum densities scale like $R^{-2}$ (Eqn. 2). At large radii, much higher densities are still needed to moderate the ionizations compared to what is expected from the simple scaling $n_H\sim N_H/\Delta R_{flow}$ for continuous flows. In particular, the derived radial filling factor $\Delta R/R$ scales like $R^{-1}$. Even the recent models that place some FeLoBAL outflows at $\sim$kpc distances require small clouds with radial filling factors $\Delta R_{clouds}/R\sim 10^{-5}$ \citep{Faucher12b}. 

In the transverse direction, the outflow sizes are constrained by the size of the quasar continuum source. Mini-BAL troughs that reach $\ga$15\% below the continuum imply absorbing regions that cover $\ga$15\% of the continuum source in projected area. For the quasar luminosities in our sample, the UV continuum region at $\lambda = 1500$ \AA\ should have radius $R_{UV}\sim 10^{16}$ cm \citep[intermediate between the predictions of standard accretion disk models and the larger numbers from microlensing observations, e.g.,][]{Blackburne11}. This means that the transverse width of the mini-BAL flow is at least $2\sqrt{0.15\,R_{UV}^2\cos\theta}\, \sim\, 8\times 10^{15}$ cm, where $\theta$ is our viewing angle of the disk measured from the polar axis and the numerical result assumes $\theta\sim 0$. Altogether these estimates imply that the outflow regions are thin and wide like ``pancakes'' viewed face on, or they occupy larger volumes like a fine spray of tiny clouds with a very small volume filling factor \citep[see also][]{Gabel06,Hamann11,Rogerson11}. 

These arguments for small dense clouds exactly parallel early photoionization studies of the broad emission line regions \citep[where densities $n_H \ga 10^{10}$ \cmn\ are required, e.g.,][]{Ferland92} and studies of BAL outflows before the shielding model was introduced \citep{Junkkarinen83,Weymann85,Hamann93,Arav94,Arav94b,Turnshek95}. Small dense clouds present a major theoretical challenge to understand how they are created and maintained \citep{Murray95,deKool97}. They also pose an observational  challenge to explain the smooth appearance of broad absorption and broad emission line profiles. If the clouds individually have only thermal velocity dispersions (roughly 15 \kms\ for hydrogen in a photoionized gas with $T\sim 20,000$ K), then the cloud numbers required for smooth line profiles are estimated to be $\ga$$10^5$ in BALs \citep{Junkkarinen02} and $>$$10^6$ or $>$$10^8$ for the broad emission lines \citep{Arav97,Dietrich99}. 

The shielding model introduced by \cite{Murray95} and \cite{Murray97} was designed to avoid these problems for quasar outflows by permitting moderate ionization levels at low gas densities in smooth continuous outflow streams. 
However, the absence of significant shielding in NAL and mini-BAL outflows means that tiny dense clouds are once again required. If we accept the prevailing view that BALs, NALs, and mini-BALs are all part of the same general outflow phenomenon, based on their many observational similarities \citep[e.g.,][]{Ganguly01,Chartas09,Elvis12,Hamann12}, then we need to consider that dense substructures and small volume filling factors are a common characteristic of all quasar outflows. These substructures might resemble the small-scale clumps inferred from fluctuations in the line-of-sight covering fraction in one well-measured BAL outflow \citep{Hall07}. 

The most promising theoretical scheme to explain dense substructured outflows is probably magnetic confinement in a magnetic disk wind \citep[e.g.,][]{Konigl94,Arav94b,deKool95}. Some type of confinement seems necessary because the cloud dissipation times are less than a characteristic flow time, e.g., $R/v \sim 44$ yr for $v\sim 0.15$c and $R\sim 2$ pc. For example, in the extreme case of outflows comprised of a single ``bullet" cloud the thickness would be $\Delta R_{clouds}\la 3\times 10^{14}$ cm or $\la$$5\times 10^{15}$ cm for maximum shielding and mini-BAL regions with $N_H\la 10^{21}$ \cmN\ or $\la$$10^{22}$ \cmN , respectively, based the minimum densities in Figure 5. If the internal velocity dispersion is similar to the measured line widths, $\Delta v\sim 1500$ \kms , then the dissipation times $\Delta R_{clouds}/\Delta v$ would be only $\la$0.2\% or $\la$2\%, respectively, of the flow time given above. In the more likely event that the flows are comprised of many clouds with individual sizes $\sim$$\Delta R_{clouds}/N$, where $N$ is the number of clouds, the dissipation times cannot be longer than a sound  crossing time or $\la$10\%/$N$ or $\la$160\%/$N$ of the flow time \citep[for the same two cases of maximally shielded mini-BAL clouds with $T\sim 20,000$ K, see Eqn. 2 in][]{Faucher12b}. 

In the \cite{deKool95} model, which draws upon earlier work on quasar broad emission line regions \citep{Rees87,Emmering92}, many small clouds with a low volume filling factor are driven out by radiative forces while being confined by magnetic pressure. One advantage of magnetic confinement is that individual clouds can have super-thermal velocity dispersions and, therefore, fewer clouds are needed to explain the observed smooth line profiles \citep[see][]{Bottorff00b}.

Another advantage of magnetic confinement is that the clouds can maintain a roughly constant density and ionization across the acceleration region \citep{deKool97}. This can be essential to preserve sufficient opacities for radiative driving. In contrast, simple mass continuity in steady-state continuous flows (with volume filling factor unity and expanding into a fixed solid angle) requires 
\begin{equation}
\dot{M}_{out} \ = \ 4\pi R^2 \rho\, v \ = \ {\rm constant}
\end{equation}
where $\dot{M}_{out}$ is the mass loss rate and $\rho$ is the mass density. While too simple to describe real quasar outflows, this relationship compared to Equation 1 shows that the ionization parameter in smooth flows scales approximately with the outflow velocity, i.e., $U\propto v$. Thus the ionization conditions needed for radiative acceleration are very fragile: At the locations where flow material is just distant enough or just shielded enough to have low-enough ionizations and sufficient opacities for radiative driving, the subsequent acceleration leads to a dramatic drop in the density (Eqn. 5), a rise in the ionization parameter (Eqn. 1), and a sharp decrease in the acceleration. As a result, smooth flows can easily ``fail'' in the sense of not reaching the gravitational escape speed or not reaching speeds high enough to match observations \citep[see also][Hamann et al. in prep.]{Proga04}. It is not clear if current theoretical models with radiative driving have fully solved this problem, even with radiative shielding. It is clear, however, that the problem is avoided altogether if the outflows are composed of small dense clouds confined by external forces. 

Finally, we note that none of this discussion answers the question of why the outflows in our quasar sample reach extreme speeds $>$0.1$c$. If higher flow speeds require smaller launch radii (e.g., by factors of a few compared to normal BAL and mini-BAL outflows, \S1 and \S4.1), then what different physical conditions allow that to happen? One possibility is that extreme speeds are aided by unusually soft far-UV continua \citep[e.g.,][]{Baskin13}. This might help maintain lower degrees of ionization (favorable for radiative driving) closer to the central SMBH. A quick inspection of the SDSS spectra indicates that the quasars in our sample have weak or absent HeII \lam 1640 emission, consistent with softer far-UV continua and higher outflow speeds observed in BALs \citep[][and A. Baskin, private comm.]{Baskin13}. However, reasonable changes to the emitted far-UV spectra cannot solve the over-ionization problem discussed above. Small dense flow structures are still needed if there is not substantial shielding.

\section{Summary}

We describe rest-frame UV and X-ray spectra of 8 quasars with mini-BAL outflows at extreme speeds in the range 0.1c to 0.2c. We constrain basic outflow properties and test the hypothesis that extreme speeds require a strong radiative shield for extreme radiative acceleration. Our main results are the following:

1) At least 5 of the 7 bona fide mini-BALs (\S2) varied significantly between observations separated by $\sim$2-3 years in the quasar rest frame (\S4.1). In one well-studied case, the variability time is $\la$1 yr \citep{Paola11}. These results are consistent with BAL variability studies and, in particular, with an observed trend for greater variability at higher speeds \citep[e.g.,][]{Capellupo11}.

2) We interpret the mini-BAL variations in terms of outflow clouds crossing our lines of sight (\S4.1). This implies radial distances $R\la 10$ pc from the central SMBH for variability times $\la$3 yr, or $R\la 1$ pc for times $\la$1 yr \citep{Capellupo12}. 

3) The X-ray absorption is typically weak or moderate, with total neutral-equivalent column densities, $N_H \la {\rm few}\times 10^{22}$ \cmN , consistent with previous studies of mini-BAL and NAL outflows but substantially less than BAL quasars (\S4.2). Thus we find no evidence for strong radiative shielding related to the extreme outflow speeds in our mini-BAL sample. 

4) Cloudy photoionization models show that the maximum amounts of shielding consistent with our data are not sufficient to control the outflow ionizations (\S5) and, therefore, not important for the acceleration (\S6). These results apply to shields that fully cover the emission source at all wavelengths. However, shielding in more complex geometries seems unlikely for NAL and mini-BAL outflows because the alleged shield would need tuning to cover the far-UV emission source (essential for shielding) while {\it not} blocking the X-ray emission near the central SMBH and {\it not} covering the near-UV source farther out in the accretion disk (\S5.3).

5) We propose that the outflow ionizations are maintained at moderate levels (low enough for \civ\ and \ovi\ absorption lines) by high gas densities in small outflow substructures (\S6). If the mini-BAL gas is at $R\sim 2$ pc with column density $N_H\la 10^{21}$ \cmN\ behind a maximum shield, then the cloud densities must be $n_H\ga 4\times 10^6$ \cmn\  corresponding to total radial extents $\Delta R_{clouds} \la 3\times 10^{14}$ cm and filling factors $\Delta R_{clouds}/R \la 5\times 10^{-5}$. If the shielding is negligible, then the cloud properties are $n_H\ga 3\times 10^7$ \cmn , $\Delta R_{clouds} \la 3\times 10^{13}$ cm,  and $\Delta R_{clouds}/R \la 5\times 10^{-6}$. Compared to transverse absorber sizes $\ga$$8\times 10^{15}$ cm (based on measured line depths), the outflows have geometries like thin ``pancakes" viewed face on or they occupy much larger volumes like a mist of many dense clouds with a very small volume filling factor.

6) In the context of popular models that have BALs, mini-BALs, and (some) outflow NALs forming in the same general outflow phenomenon, our results suggest that all types of quasar accretion disk outflows are composed of small dense substructures and that radiative shielding is not important for the acceleration. The best theoretical scheme to explain substructured flows is probably with magnetic confinement in magnetic disk winds \citep[e.g.,][]{deKool95,Emmering92}.

\section*{Acknowledgments}

We are grateful to Alexei Baskin, Gary Ferland, and Ari Laor for helpful discussions, and to Gary Ferland for making the Cloudy photoionization code available as a public resource. We thank the staffs of the Chandra and MDM Observatories for their willing help with the observations. This work was supported by NASA through the Smithsonian Astrophysical Observatory award GO1-12146A/B/C. FH also acknowledges support from the USA National Science Foundation grant AST-1009628. ME and JC acknowledge support from NSF grant AST-0807993.



\label{lastpage}

\end{document}